\documentclass[aps,prd,preprint,superscriptaddress,showpacs,floatfix,nofootinbib]{revtex4}

\usepackage{graphicx}

\newcommand{\bmat}{\left(\begin{array}}
\newcommand{\emat}{\end{array}\right)}
\newcommand{\beq}{\begin{equation}}
\newcommand{\eeq}{\end{equation}}

\def\yzero{\smash{\hbox{$y\kern-4pt\raise1pt\hbox{${}^\circ$}$}}}

\def\-{\hphantom{-}}

\def\s2{\frac{1}{\sqrt2}}

\def\beq{\begin{equation}}
\def\eeq{\end{equation}}
\def\beqa{\begin{eqnarray}}
\def\eeqa{\end{eqnarray}}

\def\IF{\relax{\rm I\kern-.18em F}}
\def\II{\relax{\rm I\kern-.18em I}}
\def\IP{\relax{\rm I\kern-.18em P}}

\def\Dsl{\,\raise.15ex\hbox{/}\mkern-13.5mu D} 

\def\IC{\bf C}
\def\IZ{\bf Z}

\def\z2z2{$\IC^3/(\IZ_2\times\IZ_2)$}

\def\s{\sigma}

\def\z{\zeta}

\def\bo{{\raise-.3ex\hbox{\large$\Box$}}}               
\def\face{{\raise.2ex\hbox{$\displaystyle \bigodot$}\mskip-2.2mu \llap {$\ddot
        \smile$}}}                                      

\def\leftrightarrowfill{$\mathsurround=0pt \mathord\leftarrow \mkern-6mu
        \cleaders\hbox{$\mkern-2mu \mathord- \mkern-2mu$}\hfill
        \mkern-6mu \mathord\rightarrow$}       
\def\dvec#1{\vbox{\ialign{##\crcr
        \leftrightarrowfill\crcr\noalign{\kern-1pt\nointerlineskip}
        $\hfil\displaystyle{#1}\hfil$\crcr}}}           

\def\beq{\begin{equation}}
\def\eeq{\end{equation}}

\def\beqx{\begin{displaymath}}
\def\eeqx{\end{displaymath}}

\def\beqa{\begin{eqnarray}}
\def\eeqa{\end{eqnarray}}

\begin{document}

\title{ Flux Parameter Spaces in Type II Vacua}

\author{Ching-Ming Chen}

\affiliation{George P. and Cynthia W. Mitchell Institute for
Fundamental Physics, Texas A$\&$M University, College Station, TX
77843, USA }

\author{Tianjun Li}

\affiliation{George P. and Cynthia W. Mitchell Institute for
Fundamental Physics, Texas A$\&$M University, College Station, TX
77843, USA }

\affiliation{
Institute of Theoretical Physics, Chinese Academy of Sciences,
 Beijing 100080, China
}

\author{Dimitri V. Nanopoulos}

\affiliation{George P. and Cynthia W. Mitchell Institute for
Fundamental Physics, Texas A$\&$M University, College Station, TX
77843, USA }

\affiliation{Astroparticle Physics Group, Houston Advanced
Research Center (HARC), Mitchell Campus, Woodlands, TX 77381, USA}

\affiliation{Academy of Athens, Division of Natural Sciences, 28
Panepistimiou Avenue, Athens 10679, Greece }

\date{\today}

\begin{abstract}

We study the flux parameter spaces for semi-realistic 
supersymmetric Pati-Salam models in the AdS vacua on 
Type IIA orientifold and realistic supersymmetric 
Pati-Salam models in the Minkowski vacua on Type IIB 
orientifold. Because the fluxes can be very large, 
we show explicitly that there indeed exists a huge 
number of semi-realistic Type IIA and realistic Type 
IIB flux models. In the Type IIA flux models, in the 
very large flux limit, the theory can become weakly 
coupled and the AdS vacua can approach to the Minkowski 
vacua. In a series of realistic Type IIB flux models, 
at the string scale, the gauge symmetry can be broken 
down to the Standard Model (SM) gauge symmetry, the gauge 
coupling unification can be achieved naturally, all the 
extra chiral exotic particles can be decoupled, and the 
observed SM fermion masses and mixings can be obtained 
as well. In particular, the real parts of the dilaton, 
K\"ahler moduli, and the unified gauge coupling are 
independent of the very large fluxes. The very large 
fluxes only affect the real and/or imaginary parts of 
the complex structure moduli, and/or the imaginary parts 
of the dilaton and K\"ahler moduli. However, these 
semi-realistic Type IIA and realistic Type IIB flux 
models can not be populated in the string landscape.

\end{abstract}

\pacs{11.10.Kk, 11.25.Mj, 11.25.-w, 12.60.Jv}

\preprint{ACT-02-08, MIFP-08-11}

\maketitle


\section{Introduction}

One of the  most challenging problems in string theory is the
construction of realistic string vacua, which can give
us the low-energy supersymmetric Standard Model (SM) and
stabilize the moduli fields. Such constructions will give us a
bridge between string theory and low-energy  realistic
particle physics. With M-theory, we can probe 
the physical string vacua not only in  
perturbative heterotic string theory, but also in 
perturbative Type I, Type IIA and Type IIB superstring theory.
Especially, because of the advent of D-branes, we can
construct consistent four-dimensional $N=1$ supersymmetric  chiral 
models with non-Abelian gauge symmetry on Type II orientifolds, 
by employing conformal field theory techniques in the open string 
sector~\cite{Polchinski:1995df}.

During the last few years, Type II orientifolds with intersecting
D-branes have been highly interesting in the string model building
where the chiral fermions come from the intersections of the D-branes
in the internal space~\cite{Berkooz:1996km}
with T-dual decription in terms of magnetized D-branes~\cite{Bachas:1995ik}. 
On Type IIA orientifolds with intersecting D6-branes, 
a large number of non-supersymmetric three-family Standard-like models 
and Grand Unified Theories (GUTs), which satisfy the Ramond-Ramond (RR)
tadpole cancellation conditions, were 
constructed~\cite{Blumenhagen:2000wh, Angelantonj:2000hi, 
Ibanez:2001nd, Blumenhagen:2005mu}. However, there
generically exist the uncancelled Neveu-Schwarz-Neveu-Schwarz (NSNS)
tadpoles and the gauge hierarchy problem. To solve these problems,
the semi-realistic supersymmetric standard-like models and GUTs 
have been constructed  in Type IIA theory on $T^6/(\IZ_2\times \IZ_2)$ 
orientifold~\cite{CSU1, CSU2, CP, Cvetic:2002pj, CLL, Cvetic:2004nk,
Chen:2005ab, Chen:2005mj} and other orientifolds~\cite{ListSUSYOthers}.
In particular, only the Pati-Salam models can give us all
the SM fermion Yukawa couplings at the stringy tree level, 
and we can explain the SM fermion masses
and mixings in one model~\cite{Chen:2007px, Chen:2007zu}.

Although some of the complex structure moduli
(in Type IIA picture) and the dilaton field might be stabilized
due to the gaugino condensation in the hidden sector in 
some models (for example, see Ref.~\cite{CLW}), 
the moduli stabilization is 
still a big challenge.  Recently, important progresses
have been made by introducing background fluxes. In Type IIB theory, 
the RR fluxes and NSNS fluxes generate a superpotential~\cite{GVW} 
that depends on the dilaton and complex structure moduli, 
and then stabilize
these fields dynamically~\cite{Giddings:2001yu, Kachru:2002sk}. 
With non-perturbative effects, one
can further determine the K\"ahler moduli~\cite{Kachru:2003aw}. 
For such kind of model building, the RR and NSNS fluxes contribute to 
large positive D3-brane charges due to the Dirac 
quantization~\cite{CU, BLT}. Thus, they modify the global RR tadpole
cancellation conditions significantly and impose strong
constraints on the consistent model 
building~\cite{MS, CL, Cvetic:2005bn, Kumar:2005hf, Chen:2005cf}. 
In addition, turning on the RR, NSNS, metric, and non-geometric 
fluxes~\cite{Shelton:2005cf, Aldazabal:2006up, Villadoro:2006ia}, 
we can stabilize close string moduli in supersymmetric 
Minkowski vacua~\cite{Chen:2007af}. In particular, 
these fluxes can contribute 
negative D-brane charges to the RR tadpole cancellation conditions,
and then we can relax the RR tadpole cancellation conditions 
elegantly~\cite{Chen:2007af}.  
In Type IIA theory with RR, NSNS, and metric 
fluxes~\cite{Grimm:2004ua, Villadoro:2005cu, Camara:2005dc}, we can
 stabilize the moduli in supersymmetric AdS vacua as well 
as relax the RR tadpole cancellation 
conditions~\cite{Camara:2005dc, Chen:2006gd}. 
Interestingly, by relaxing the RR tadpole cancellation conditions,
we can construct semi-realistic 
Type IIA~\cite{Chen:2006gd, Chen:2006ip}
and realistic IIB~\cite{Chen:2007af} Pati-Salam flux models 
that can explain the SM fermion masses and mixings.

One of the most interesting consequences from Type II flux 
compactifications is that there may exist a large number of 
  meta-stable vacua~\cite{Susskind:2003kw, Denef:2004ze}. 
The ensemble of these vacua is 
called string landscape~\cite{Susskind:2003kw}.  
With the ``weak anthropic principle''~\cite{Weinberg:1987dv}, 
this proposal may provide the first concrete explanation of
the very tiny value of the cosmological constant
which can take only the discrete values.
Because the fluxes can contribute negative D-brane charges to 
the RR tadpole cancellation conditions
 in Type IIA and IIB theories with general flux 
compactifications~\cite{Chen:2007af, Camara:2005dc, Chen:2006gd},
we can easily construct many (or even infinite) 
Type IIA and IIB flux vacua by increasing the corresponding fluxes. 
Therefore, the remain interesting open questions are whether we can have 
a huge number of semi-realistic Type IIA and
 realistic Type IIB flux vacua, and whether
these  vacua can be populated in the string landscape.

In this paper, we shall study the flux parameter spaces
in details for semi-realistic supersymmetric Pati-Salam 
models in the AdS vacua on  Type IIA orientifold
and realistic supersymmetric Pati-Salam 
models in the Minkowski vacua on Type IIB orientifold
with general flux compactifications. We
show explicitly that there indeed exists a huge number of
semi-realistic Type IIA and  realistic Type IIB Pati-Salam flux models. 
These discussions also confirm the possibility of the string landscape. 
However, it seems to us that these semi-realistic  Type IIA
and realistic Type IIB flux models can not be populated
in the string landscape. The point is the following:
we can easily construct much more models that do not have the SM 
gauge symmetry  while satisfy the same model building constraints,
for concrete examples, see the statistical study in 
Ref.~\cite{Gmeiner:2005vz}.

First, we briefly review the supersymmetric intersecting D6-brane 
model building in the AdS vacua on Type IIA orientifold with flux 
compactifications. In general, even though we take the RR fluxes 
$e$ and/or $e_0$ (for detail definitions, see the following 
discussions and those in Ref.~\cite{Camara:2005dc}) 
to be very large, we do not change
 the gauge groups and the particle spectra of the D6-brane 
models since the fluxes $e$ and $e_0$ do not 
contribute to the RR tadpole cancellation 
conditions. In particular,  if we take both $e$ and $e_0$ to 
be very large while keeping $(e_0a-eh_0)$ as a constant where $a$ and $h_0$ 
are respectively the metric and NSNS fluxes, the very large 
fluxes $e$ and $e_0$ will not affect the dilaton $s$, K\"ahler moduli 
$t_i$, complex structure moduli $u_i$, and the cosmological constant 
of the models. In this case, we can take $e$ and $e_0$ to be very large, 
and keep the same main properties of the models, for example, the gauge 
symmetries, the gauge coupling constants, the particle spectra, 
 and the cosmological constant, etc.
The only difference is that one linear combination of imagininary
parts of $s$ and $u_i$ ($ 3a {\rm Im} s +\sum_{i=1}^3 b_i {\rm Im} u_i$)
will be very large and proportional to $e$ where $b_i$ are metric
fluxes. Moreover, if $(e_0a-eh_0)$
is very large, we show that the theory will be very weakly coupled, and 
the magnitude of the cosmological constant will be very small 
and proportional to $(e_0a-eh_0)^{-5/3}$ so that the AdS vacua
approach to the Minkowski vacua. And if only $e_0$ is very large, 
($ 3a {\rm Im} s +\sum_{i=1}^3 b_i {\rm Im} u_i$) will be a constant
as well since it only depends on $e$. Next, we take the other fluxes 
$a$, $h_0$, $m$, and/or $q$ to be very large where $m$ and $q$ are RR fluxes. 
Interestingly, only the quadratic combination $(h_0m +3qa)$ appears
in the RR tadpole cancellation conditions. And if $(h_0m +3qa)<0$,
the fluxes contribute negative D-brane charges to all
the RR tadpole cancellation conditions. Thus, when $a$, $h_0$, $m$, and/or $q$
are very large, we consider two kinds of models:
(1) $(h_0m +3qa)$ is a constant; 
(2) $(h_0m +3qa)$ is very large.  
In the first kind of Pati-Salam models which was constructed and 
studied by us previously~\cite{Chen:2006gd}, 
we take $h_0$, $m$, and $q$ to be very large while
keeping $(h_0m +3qa)$ as a constant. Thus, we do not change 
 the RR tadpole concellation conditions, the 
gauge symmetries, and the particle spectra 
of the models. In the very large flux limit,
the theory becomes very weakly coupled, 
the magnitude of the cosmological constant will become very small,
and then the AdS vacua also approach to the Minkowski vacua. 
In the second kind of Pati-Salam models, 
we not only take $a$ and $h_0$ to be very large, but also take
$(h_0m +3qa)$ to be very large. If $(h_0m +3qa)$ is negative 
and very large,
we can always satisfy the RR tadpole cancellation conditions
by introducing more D-branes. However, we will have many
chiral exotic particles that can not be decoupled easily. Thus,
we take the complex structure moduli to be very large so that 
only one of the RR tadpole cancellation conditions
will become very large and proportional to $a$ or $h_0$.
We show that at the string scale,
the gauge symmetry can be broken down to the SM gauge
symmetry, and the exotic particles might be decoupled.
The gauge couplings for $SU(2)_L$ and $SU(2)_R$ are half of these
for $SU(3)_C$ and $U(1)_{B-L}$. In the very large flux limit,
the gauge coulings approach to the fixed constants, but
the magnitude of the cosmological constant  will be very large which
can be made very small again if we introduce very large $e$ and/or $e_0$ 
fluxes.

Second, we briefly review the supersymmetric D-brane model building 
in the Minkowski vacua on Type IIB orientifold with general
flux compactifications. We construct a series of realistic
Pati-Salam models with  gauge groups 
$U(4)_C \times U(2)_L \times U(2)_R$ and $USp(10) \times USp(6(\kappa-1))^3$
in the observable sector and hidden sector, respectively, where
$\kappa=1, 2, 3, ...$.  At the string scale, the gauge symmetry can be 
broken down to the SM gauge symmetry, the gauge coupling 
unification can be achieved naturally, and all the extra chiral exotic 
particles can be decoupled so that we have the supersymmetric SMs with/without 
SM singlet(s) below the string scale. The observed SM fermion masses and mixings 
can also be obtained. In addition, the unified gauge coupling, the dilaton, 
the complex structure moduli, the real parts of the K\"ahler moduli and 
the sum of the imaginary parts of the K\"ahler moduli can be determined 
as functions of the four-dimensional dilaton and fluxes, and can be 
estimated as well. In particular, in the
 very large flux limit, we can choose 
the unified gauge coupling at the string scale as the unified gauge 
coupling in the supersymmetric SMs at the GUT scale. 
In other words, we can choose the real parts of the dilaton ($s$) and 
K\"ahler moduli ($t_i$) as constants because they are
independent of the fluxes. 
And we emphasize that the cosmological constant is always zero in these models
due to the Minkowski vacua. Moreover, we consider the flux parameter spaces
in two kinds of the models: $\kappa=1$ and $\kappa > 1$ . For the 
first kind of models with
 $\kappa=1$, we do not have gauge symmetries $USp(6(\kappa-1))^3$,
and we have constructed and studied such kind of models 
previously~\cite{Chen:2007af}. 
In these models, the very large fluxes do not contribute to the RR tadpole
cancellation conditions, and we find two kinds of solutions to 
the flux consistent equations. In the very large flux limit, 
the real part of the complex 
structure moduli ($u_1=u_2=u_3\equiv u$) and the sum of the imaginary 
parts of the K\"ahler moduli ($t\equiv t_1+t_2+t_3$)
are constants and independent of the very large fluxes. Only the imaginary
parts of the dilaton and complex structure moduli will be very large
and proportional to the very large fluxes.
For the second kind of models with $\kappa > 1$, we obtain four kinds of 
solutions to the flux consistent equations, and show that there
indeed exist a huge number of the realistic flux vacua.
In the very large $\kappa$ limit, we consider the asymptotic behaviour
 for ${\rm Re} u$, ${\rm Im}s$, ${\rm Im}t$, and ${\rm Im}u$,
and present some concrete examples.

This paper is organized as follows. In Section II, we briefly
review the supersymmetric intersecting D6-brane model building 
in the AdS vacua on Type IIA orientifold with flux compactifications
and the general case with very large fluxes $e$ and/or $e_0$.
And we study two kinds of the Type IIA Pati-Salam flux models in Section III.
In Section IV, we briefly review the supersymmetric D-brane 
model building in the Minkowski vacua on Type IIB orientifold with 
general flux compactifications.
We study a series of realistic
Pati-Salam flux models in Section V.
Discussion and conclusions are given in Section VI.



\section{Flux Model Building on Type IIA orientifold}

Let us briefly review the rules for the
intersecting D6-brane model building in Type IIA theory on
$\mathbf{T^6}$ orientifold with flux 
compactifications~\cite{Villadoro:2005cu,Camara:2005dc}.
 We consider $\mathbf{T^6}$ to be a six-torus factorized as 
$\mathbf{T^6} = \mathbf{T^2} \times \mathbf{T^2} \times \mathbf{T^2}$
whose complex coordinates are $z_i$, $i=1,\; 2,\; 3$ for the
$i$-th two-torus, respectively. 
We implement an orientifold projection $\Omega R$, where $\Omega$
is the world-sheet parity, and $R$ acts on the complex coordinates as
\begin{equation}
R:(z_1,z_2,z_3)\rightarrow(\overline{z}_1,\overline{z}_2,\overline{z}_3)~.~\,
\end{equation}
Thus, we have the orientifold 6-planes (O6-planes) under
the actions of $\Omega R$.
In addition, we introduce some stacks of D6-branes
 which wrap on the factorized three-cycles.
There are two kinds of complex structures
consistent with orientifold projection for a two-torus --
rectangular and tilted~\cite{CSU2,LUII}. If we denote the
homology classes of the three cycles wrapped by $a$ stack of $N_a$ D6-branes
 as $n_a^i[a_i]+m_a^i[b_i]$ and $n_a^i[a'_i]+\tilde{m}_a^i[b_i]$
with $[a_i']=[a_i]+\frac{1}{2}[b_i]$ for the rectangular and
tilted two-tori respectively, we can label a generic one cycle by
$(n_a^i,l_a^i)$ in which
 $l_{a}^{i}\equiv m_{a}^{i}$ for a rectangular two-torus
while $l_{a}^{i}\equiv 2m_{a}^{i}=2\tilde{m}_{a}^{i}+n_{a}^{i}$ for
a tilted two-torus~\cite{Cvetic:2002pj}.
For $a$ stack of $N_a$ D6-branes along
the cycle $(n_a^i,l_a^i)$, we also need to include their $\Omega
R$ images $N_{a'}$ with wrapping numbers $(n_a^i,-l_a^i)$. For the
D6-branes on the top of O6-planes, we count them and their $\Omega R$
images independently. So, the homology three-cycles for $a$ stack
of $N_a$ D6-branes and its orientifold image $a'$ are
\begin{eqnarray}
[\Pi_a]=\prod_{i=1}^{3}\left(n_{a}^{i}[a_i]+2^{-\beta_i}l_{a}^{i}[b_i]\right),\;\;\;
\left[\Pi_{a'}\right]=\prod_{i=1}^{3}
\left(n_{a}^{i}[a_i]-2^{-\beta_i}l_{a}^{i}[b_i]\right)~,~\,
\end{eqnarray}
where $\beta_i=0$ if the $i$-th two-torus is rectangular and
$\beta_i=1$ if it is tilted. And the homology three-cycle wrapped
by the  O6-planes is
\begin{eqnarray}
\Omega R: [\Pi_{O6}]= 2^3
[a_1]\times[a_2]\times[a_3]~.~\,
\end{eqnarray}
Therefore, the intersection numbers are
\begin{eqnarray}
I_{ab}=[\Pi_a][\Pi_b]=2^{-k}\prod_{i=1}^3(n_a^il_b^i-n_b^il_a^i)~,~\,
\end{eqnarray}
\begin{eqnarray}
I_{ab'}=[\Pi_a]\left[\Pi_{b'}\right]=-2^{-k}\prod_{i=1}^3(n_{a}^il_b^i+n_b^il_a^i)~,~\,
\end{eqnarray}
\begin{eqnarray}
I_{aa'}=[\Pi_a]\left[\Pi_{a'}\right]=-2^{3-k}\prod_{i=1}^3(n_a^il_a^i)~,~\,
\end{eqnarray}
\begin{eqnarray}
 I_{aO6}=[\Pi_a][\Pi_{O6}]=-2^{3-k} l_a^1l_a^2l_a^3~,~\,
\label{intersections}
\end{eqnarray}
 where  $k=\beta_1+\beta_2+\beta_3$ is
the total number of tilted two-tori.

\begin{table}[h]
\renewcommand{\arraystretch}{1.5}
\center
\begin{tabular}{|c||c|}
\hline

Sector & Representation   \\ \hline \hline

$aa$ & $U(N_a )$ vector multiplet and 3 adjoint chiral multiplets \\
\hline

$ab+ba$ & $I_{ab}$ $ (N_a,
\overline{N_b})$ chiral multiplets  \\ \hline

$ab'+b'a$ & $I_{ab'}$ $ (N_a,
N_b)$ chiral multiplets  \\ \hline

$aa'+a'a$ & $  \frac 12 (I_{aa'} + I_{aO6}) $
anti-symmetric chiral multiplets  \\

  & $\frac 12 (I_{aa'} - I_{aO6})$
symmetric chiral multiplets  \\ \hline

\hline

\end{tabular}
\caption{The general spectrum for the intersecting D6-brane model
building in Type IIA theory on $\mathbf{T^6}$ orientifold
with flux compactifications.}
\label{Spectrum-T6}
\end{table}

For $a$ stack of $N_a$ D6-branes whose homology three-cycles
are not invariant under $\Omega R$, we have
$U(N_a)$ gauge symmetry. Otherwise, we obtain $USp(2N_a)$ 
gauge symmetry. The general spectrum of D6-branes' intersecting 
at generic angles, which is valid for both rectangular and 
tilted two-tori, is given in Table \ref{Spectrum-T6}. 
The four-dimensional $N=1$
supersymmetric models on Type IIA orientifolds with intersecting
D6-branes are mainly constrained in two aspects:
four-dimensional $N=1$  supersymmetry conditions, and
RR tadpole cancellation conditions.

To simplify the notation, we define the following products of wrapping
numbers
\begin{eqnarray}
\begin{array}{rrrr}
A_a \equiv -n_a^1n_a^2n_a^3~, & B_a \equiv n_a^1l_a^2l_a^3~,
& C_a \equiv l_a^1n_a^2l_a^3~, & D_a \equiv l_a^1l_a^2n_a^3~, \\
\tilde{A}_a \equiv -l_a^1l_a^2l_a^3~, & \tilde{B}_a \equiv
l_a^1n_a^2n_a^3~, & \tilde{C}_a \equiv n_a^1l_a^2n_a^3~, &
\tilde{D}_a \equiv n_a^1n_a^2l_a^3~.\,
\end{array}
\label{variables}
\end{eqnarray}

(1) {\it Four-Dimensional $N=1$  Supersymmetry Conditions}

The four-dimensional $N=1$ supersymmetry  can be preserved by the
orientation projection ($\Omega R$) if and only if the rotation angle of any
D6-brane with respect to any O6-plane is an element of
$SU(3)$~\cite{Berkooz:1996km}, {\it i.~e.},
$\theta_1+\theta_2+\theta_3=0 $ mod $2\pi$, where $\theta_i$ is
the angle between the $D6$-brane and the O6-plane on the
$i$-th two-torus. Then the supersymmetry conditions can be
rewritten as~\cite{Cvetic:2002pj}
\begin{eqnarray}
x_A\tilde{A}_a+x_B\tilde{B}_a+x_C\tilde{C}_a+x_D\tilde{D}_a~=~0~,~\,
\label{susyconditions}
\end{eqnarray}
\begin{eqnarray}
 A_a/x_A+B_a/x_B+C_a/x_C+D_a/x_D~<~0~,~\,
\label{susyconditionsII}
\end{eqnarray}
where $x_A=\lambda,\;
x_B=\lambda 2^{\beta_2+\beta3}/\chi_2\chi_3,\; x_C=\lambda
2^{\beta_1+\beta3}/\chi_1\chi_3,$ and $x_D=\lambda
2^{\beta_1+\beta2}/\chi_1\chi_2$ in which $\chi_i=R^2_i/R^1_i$ are the
complex structure parameters and $\lambda$ is a positive real number.

(2) {\it RR Tadpole Cancellation Conditions}

 The total RR charges from the D6-branes and O6-planes and from the metric, NSNS,
and RR  fluxes must vanish since the RR field flux lines are conserved.
With the filler branes on the top of the O6-plane, we obtain the RR
tadpole cancellation conditions~\cite{Villadoro:2005cu,Camara:2005dc}:
\begin{eqnarray}
2^k N_{O6} - \sum_a N_a A_a + {1\over 2}(m h_0  + q_1 a_1 + q_2 a_2 + q_3
a_3)& = & 16 ~,~\,
\label{tadodhz2z2I}
\end{eqnarray}
\begin{eqnarray}
 \sum_a 2^{-\beta_2-\beta3} N_a B_a
+ {1\over 2} (m h_1 - q_1 b_{11} - q_2 b_{21} -
q_3 b_{31}) & = & 0 ~,~\,
\label{tadodhz2z2II}
\end{eqnarray}
\begin{eqnarray}
\sum_a 2^{-\beta_1-\beta_3} N_a C_a
+ {1\over 2} ( m h_2 - q_1 b_{12} - q_2 b_{22} - q_3 b_{32}) & = & 0 ~,~\,
\label{tadodhz2z2III}
\end{eqnarray}
\begin{eqnarray}
\sum_a 2^{-\beta_1-\beta_2} N_a D_a + {1\over 2} (m h_3 - q_1 b_{13} - q_2 b_{23} -
q_3 b_{33}) & = & 0 ~,~\,
\label{tadodhz2z2IV}
\end{eqnarray}
where $2 N_{O6}$ are the number of filler branes wrapping along
the  O6-plane.
In addition, $a_i$ and $b_{ij}$ arise from the metric
fluxes, $h_0$ and $h_i$ arise from the NSNS fluxes, and $m$ and
$q_i$ arise from the RR fluxes. We consider these
fluxes ($a_i$, $b_{ij}$, $h_0$, $h_i$, $m$ and $q_i$)
quantized in units of 2 so that we can avoid
the subtle problems with flux Dirac quantization conditions.

In addition to the above RR tadpole cancellation conditions,
the discrete D-brane RR
charges classified by $\mathbf{Z_2}$ K-theory groups in the
presence of orientifolds, which are subtle and invisible by the
ordinary homology~\cite{MS,Witten9810188, MarchesanoBuznego:2003hp},
should also be taken into account~\cite{CU}. The K-theory conditions for a
$\mathbf{Z_2\times Z_2}$ orientifold are
\begin{equation}
\sum_a  \tilde{A}_a = \sum_a  \tilde{B}_a
  = \sum_a  \tilde{C}_a  = \sum_a  \tilde{D}_a  =
0 \textrm{ mod } 2 ~.~\,
\label{K-charges}
\end{equation}

Moreover, the Freed-Witten anomaly~\cite{Freed:1999vc} 
cancellation condition is~\cite{Camara:2005dc}
\begin{eqnarray}
-2^{-k} h_0 \tilde{A}_a + 2^{-\beta_1} h_1 \tilde{B}_a
+ 2^{-\beta_2} h_2 \tilde{C}_a
+ 2^{-\beta_3} h_3 \tilde{D}_a =0~.~\,
\end{eqnarray}

Furthermore, there are seven moduli fields in the supergravity 
theory basis, the dilaton $s$, three K\"ahler moduli $t_i$,
and three complex structure moduli $u_i$. And their real
parts are
\begin{eqnarray}
{\rm Re}s~\equiv~{{e^{-\phi_4}}\over {\sqrt {\chi_1 \chi_2 \chi_3}}}~,~
{\rm Re}t_i~\equiv~{{A_i}\over {\alpha'}}~,~
{\rm Re}u_i~\equiv~e^{-\phi_4} {\sqrt {{\chi_j \chi_k}\over {\chi_i}}}~,~\,
\end{eqnarray}
where $\phi_4$ is the four-dimensional T-duality invariant dilaton, 
$A_i$ is the area for the $i$-th two-torus, $\alpha'$ is string tension,
 and $i \not= j \not= k \not= i$. Moreover, $\phi_4$ is related
to the ten-dimensional dilaton $\phi$ as following
\begin{eqnarray}
e^{\phi_4} ~=~ {{e^{\phi}}\over {\sqrt {{\rm Re}t_1 {\rm Re}t_2  {\rm Re}t_3}}}~.~\,
\end{eqnarray}
And the four-dimensional Planck scale $M_{\rm Pl}$ is
\begin{eqnarray}
M_{\rm Pl} ~=~ {{e^{-2\phi_4}}\over {\pi \alpha'}}~.~\,
\end{eqnarray}

Moreover, the full superpotential is~\cite{Camara:2005dc}
\begin{eqnarray}
{\cal W} &=& e_0 + i h_0 S + \sum_{i=1}^3[(i e_i -a_i s -b_{ii} u_i
-\sum_{j\not= i} b_{ij} u_j)t_i -i h_i u_i] 
\nonumber\\&& -  q_1 t_2 t_3 -q_2
t_1 t_3 -q_3 t_1 t_2 + i m t_1 t_2 t_3 ~,~\,
\end{eqnarray}
where $e_0$ and $e_i$ are RR fluxes.
The K\"ahler potential for these moduli is
\begin{eqnarray}
{\cal K} = -{\rm ln}(s+{\bar s})-\sum_{i=1}^3 {\rm ln} (t_i + {\bar t}_i)
-\sum_{i=1}^3 {\rm ln} (u_i + {\bar u}_i)~.~\,
\end{eqnarray}
And the holomorphic gauge kinetic function for a generic stack of
D6-branes  is given by~\cite{Cremades:2002te, Lust:2004cx}
\begin{eqnarray}
f_a &=& {1\over {\kappa_a}}\left(  n_a^1\,n_a^2\,n_a^3\,s-
n_a^1\,m_a^2\,m_a^3\,u_1 \right.\nonumber\\&& \left.
-n_a^2\,m_a^1\,m_a^3\,u_2 -n_a^3\,m_a^1\,m_a^2\,u_3\right)~,~\,
\label{EQ-GKF}
\end{eqnarray}
where $\kappa_a$ is equal to  1 and 2 for $U(n)$ and $USp(2n)$,
respectively. We emphasize that 
$m_{a}^{i}$ is equal to $l_{a}^{i}$ and $l_{a}^{i}/2$
for  the rectangular and
tilted two-torus, respectively.

In addition, the supergravity scalar potential is
\begin{eqnarray}
V = e^{\cal K} \left({\cal K}^{i {\bar j}} D_i {\cal W} 
D_{\bar j} {\cal W} -3|{\cal W}|^2 \right)~,~\,
\end{eqnarray}
where ${\cal K}^{i {\bar j}}$ is the inverse metric of 
${\cal K}_{i {\bar j}}\equiv \partial_i \partial_{\bar j} {\cal K}$,
 $D_i {\cal W} =\partial_i {\cal W} + (\partial_i {\cal K}) {\cal W}$,
and $\partial_i = \partial_{\phi_i}$ where $\phi_i$ can be $s$, $t_i$,
and $u_i$.

In this paper, we concentrate on the supersymmetric AdS 
vacua~\cite{Camara:2005dc}.
For simplicity, we assume that the K\"ahler moduli $t_i$
satisfy $t_1=t_2=t_3$, then we obtain $q_1=q_2=q_3\equiv q$ and
$e_1=e_2=e_3\equiv e$ at the AdS vacua. To satisfy the Jacobi identities
for metric fluxes, we consider the solution $a_i=a$,
$b_{ii}=-b_i$, and $b_{ji}=b_i$ in which $j\not= i$~\cite{Camara:2005dc}.
Then, the superpotential becomes
\begin{eqnarray}
{\cal W} &=& e_0 + 3i e t -3q t^2 +im t^3 + ih_0 s -3ast
-\sum_{k=0}^3 (ih_k + b_k t) u_k~.~\,
\end{eqnarray}

For the supersymmetric AdS vacua, we have 
\begin{eqnarray}
D_{s} {\cal W} = D_{t} {\cal W} =  D_{u_i} {\cal W}=0~.~\,
\label{Eq-stu}
\end{eqnarray}
From the above equations, we obtain the following ten real 
equations for $m \not= 0$~\cite{Camara:2005dc}
\begin{eqnarray}
3a {\rm Re} s = b_i {\rm Re} u_i~,~~~{\rm for} ~i=1, 2, 3 ~,~\,
\label{fixsu}
\end{eqnarray}
\begin{eqnarray}
3h_i a + h_0 b_i = 0 ~,~~~{\rm for} ~i=1, 2, 3 ~,~\,
\label{finetune}
\end{eqnarray}
\begin{eqnarray}
3a {\rm Im} s +\sum_{i=1}^3 b_i {\rm Im} u_i = 3e - \frac{3q}{a}(3h_0
- 7a {\rm Im} t ) - \frac{3m}{a} {\rm Im} t (3h_0 - 8a {\rm Im} t)  ~,~\,
\label{flatsu}
\end{eqnarray}
\begin{eqnarray}
a {\rm Re} s = -2 {\rm Re}t (q+ m {\rm Im} t) ~,~\,
\label{fixres}
\end{eqnarray}
\begin{eqnarray}
160\lambda^3 + 186(\lambda_0-1)\lambda^2 + 27(\lambda_0-1)^2\lambda+
\lambda_0^2(\lambda_0-3) + \frac{27a^2}{m h_0^3}(e_0 a - e h_0) = 0  ~,~\,
\label{cubice}
\end{eqnarray}
\begin{eqnarray}
3a^2 ({\rm Re} t)^2= 5h_0^2 \lambda(\lambda + \lambda_0-1) ~,~\,
\end{eqnarray}
where 
\begin{eqnarray}
{\rm Im} t= (\lambda + \lambda_0)\frac{h_0}{3a} \quad , \quad 
\lambda_0=-\frac{3q a}{m h_0}  ~.~\,
\end{eqnarray}
And we require $\lambda(\lambda + \lambda_0-1) > 0$.
Also, Eq. (\ref{fixres}) can be rewritten as 
\begin{eqnarray}
3a^2 {\rm Re} s=-2 h_0 m \lambda {\rm Re} t  ~.~\,
\label{ST-relation-IIA}
\end{eqnarray}
Interestingly, it can be shown that if Eqs. (\ref{susyconditions}),
(\ref{fixsu}), and (\ref{finetune}) are satisfied, the
 Freed-Witten anomaly is automatically cancelled. Moreover, 
we obtain the cosmological constant 
\begin{eqnarray}
V_0= - \frac{a b_1 b_2 b_3 \lambda_0^2 (16 \lambda + \lambda_0 -1)}
{1920 \, q^2 \, {\rm Re} t^3 \lambda^3} ~.~\,
\label{cosmoTIIA}
\end{eqnarray}

From Eqs. (\ref{fixsu}) and (\ref{finetune}),  we obtain
\begin{eqnarray}
b_1={{3a}\over {\chi_2 \chi_3}}~,~
b_2={{3a}\over {\chi_1 \chi_3}}~,~
b_3={{3a}\over {\chi_1 \chi_2}}~,~
\label{Eq-abchi}
\end{eqnarray}
\begin{eqnarray}
h_1=-{{h_0}\over {\chi_2 \chi_3}}~,~
h_2=-{{h_0}\over {\chi_1 \chi_3}}~,~
h_3=-{{h_0}\over {\chi_1 \chi_2}}~.~
\label{Eq-hhchi}
\end{eqnarray}
Thus, the RR
tadpole cancellation conditions can be rewritten as following
\begin{eqnarray}
2^k N_{O6} - \sum_a N_a A_a + {1\over 2}(h_0 m + 3 a q)& = & 16 ~,~\,
\label{N-tadodhz2z2I}
\end{eqnarray}
\begin{eqnarray}
 \sum_a 2^{-\beta_2-\beta3} N_a B_a
- {1\over {2 \chi_2 \chi_3}} (h_0 m + 3 a q) & = & 0 ~,~\,
\label{N-tadodhz2z2II}
\end{eqnarray}
\begin{eqnarray}
\sum_a 2^{-\beta_1-\beta_3} N_a C_a
- {1\over {2 \chi_1 \chi_3}} (h_0 m + 3 a q)  & = & 0 ~,~\,
\label{N-tadodhz2z2III}
\end{eqnarray}
\begin{eqnarray}
\sum_a 2^{-\beta_1-\beta_2} N_a D_a - {1\over {2 \chi_1 \chi_2}} (h_0 m + 3 a q)
& = & 0 ~.~\,
\label{N-tadodhz2z2IV}
\end{eqnarray}
Therefore, only the flux quadratic combination $(h_0 m + 3 a q)$ 
appears in the RR tadpole cancellation condition. And
if $(h_0 m + 3 a q) < 0$, the supergravity fluxes
contribute negative D6-brane charges to all the RR tadpole
cancellation conditions, and then, the RR tadpole cancellation
conditions give no constraints on the consistent model building
because we can always introduce suitable supergravity fluxes
and some stacks of D6-branes in
the hidden sector to cancel the RR tadpoles.

In general, we can take the fluxes $e_0$ and $e$ as arbitrary 
large integers that are multiples of 2, {\it i.e.}, $e_0$ and $e$ 
can be very large. Here, for simplicity,
we assume that the fluxes $h_0$, $m$, $q$ and $a$ are the
 finite fixed integers that satisfy the
RR tadpole cancellation conditions. Interestingly,
if  both $e$ and $e_0$ are very large while  
$(e_0a-eh_0)$ is a constant, the very large 
fluxes $e$ and $e_0$ will not affect the dilaton, K\"ahler moduli, 
 complex structure moduli, and the cosmological constant 
of the models. So, we can take $e$ and $e_0$ to be very large, 
and keep the same main properties of the models, for example, the gauge 
symmetries, the gauge coupling constants, the particle spectra, 
and the cosmological constant, etc.
The only difference is that one linear combination of the imagininary
parts of $s$ and $u_i$ ($ 3a {\rm Im} s +\sum_{i=1}^3 b_i {\rm Im} u_i$)
will be very large and proportional to $e$. For the case
with very large $(e_0a-eh_0)$, we define
\begin{eqnarray}
\Delta \equiv \left[-\frac{27a^2}{160 m h_0^3}(e_0 a - e h_0)\right]^{1/3}~.~\,
\label{Delta}
\end{eqnarray}
If $e_0$ and/or $e$ are/is very large, $\Delta$ can be
very large. In particular, if we fix $e$ and only allow
$e_0$ to be very large, the 
value of $3a {\rm Im} s +\sum_{i=1}^3 b_i {\rm Im} u_i$
is fixed and will not be very large. 
In the large $\Delta$ limit, we obtain
\begin{eqnarray}
{\rm Re} t \simeq {{\sqrt {15}} \over 3} |{{h_0 \Delta}\over a}|~,~~
{\rm Re} s \simeq -{{2h_0m \Delta}\over {3a^2}} {\rm Re} t ~,~
{\rm Re} u_i \simeq -{{2h_0m \Delta}\over {a b_i}} {\rm Re} t~.~\,
\end{eqnarray}
Thus we have
\begin{eqnarray}
e^{\phi_4} \sim \Delta^{-2} ~,~~~e^{\phi} \sim \Delta^{-1/2}~,~
g^{-2}_{a, b, c} \sim \Delta^{2} ~.~\,
\end{eqnarray}
And then, the theory will be very weakly coupled.
The cosmological constant is
\begin{eqnarray}
V_0 \simeq - \frac{a b_1 b_2 b_3 \lambda_0^2 }
{120 \, q^2 \, {\rm Re} t^3 \lambda^2} \sim \Delta^{-5}~.~\,
\label{cosmoTIIA}
\end{eqnarray}
So, the magnitude of the 
cosmological constant will be very small if $\Delta$
is very large. And then the AdS vacua will approach to the
Minkowski vacua.



\section{Type IIA Flux Models}

We will present two kinds of models where some of the fluxes are free 
parameters and can be very large. In the first kind of models,  
the very large fluxes that we introduce do not change the RR tadpole
cancellation conditions. While in the second kind of models, the 
very large fluxes  do contribute to one of the D-brane RR tadpole
cancellation conditions.

\subsection{Model TIIA-I}

In all the Pati-Salam models, $SU(5)$ models, and flipped $SU(5)$ models
that have been constructed previously~\cite{Chen:2006gd, Chen:2006ip}, 
the fluxes $h_0$, $m$, $q$, and $a$ are not fixed. 
And only one quadratic combination of them ($h_0 m+3 q a$)
is determined by the RR tadpole cancellation conditions. Thus, 
three of  the four fluxes $h_0$, $m$, $q$, 
and $a$ can be free parameters. Moreover, $e$ and $e_0$ can also be taken as
 free parameters. As an example, let us study a concrete model which
is model TI-U-4 in Ref.~\cite{Chen:2006gd}. 
We present the D-brane configurations and 
intersection numbers in Table \ref{TI-U-4}, and the particle spectrum in
the observed sector in Tables \ref{TI-U-4-spectrum-A} and \ref{TI-U-4-spectrum-B}.
The phenomenological consequences have been discussed in Ref.~\cite{Chen:2006gd}.
 Similar to the discussions in Ref.~\cite{Chen:2007px, Chen:2007zu}, we can explain 
the SM fermion masses and mixings  because all the SM fermions 
and Higgs fields arise from the intersections on the third torus.



\begin{table}[h]

\begin{center}

\footnotesize

\begin{tabular}{|@{}c@{}|c||@{}c@{}c@{}c@{}||c|c||c|@{}c@{}|c|@{}c@{}||
@{}c@{}|@{}c@{}|@{}c@{}|c|@{}c@{}|} \hline

stack & $N$ & ($n_1$,$l_1$) & ($n_2$,$l_2$) & ($n_3$,$l_3$) & A &

S & $b$ & $b'$ & $c$ & $c'$ & $d$ & $d'$ & $e$ & $e'$ & $O6$

  \\ \hline \hline

$a$ & 4 & ( 0,-1) & ( 1, 1) & ( 3, 1) & 1 & -1 & 3 & 0(1) & -3 &

0(3) & 2 & 0(2) & -3 & - & 1

\\ \hline

$b$ & 2 & (-1,-1) & ( 2, 0) & (-3, 1) & 0 & 0 & - & - & 6 & 0(3) &

-1 & -5 & 6 & - & 0(1)

\\ \hline

$c$ & 2 & ( 1,-1) & (-1, 1) & ( 0,-2) & -2 & 2 & - & - & - & - &

0(10) & 2 & 0(1) & - & -2

\\ \hline \hline

$d$ & 2 & ( 2, 3) & ( 1,-1) & ( 2, 0) & 0 & 0 & - & - & - & - & -

& - & 6 & - & 0(3)

\\ \hline

$e$ & 1 & ( 1, 0) & ( 0,-2) & ( 0, 2) & 0 & 0 & - & - & - & - & -

& - & - & - & 0(4)

\\ \hline

$O6$ & 5 & ( 1, 0) & ( 2, 0) & ( 2, 0) & - & - & - & - & - & - & -

& - & - & - & -

\\ \hline

\end{tabular}
\caption{D6-brane configurations and intersection numbers
for Model TIIA-I on Type IIA $\mathbf{T^6}$ orientifold.
The complete gauge symmetry is
$[U(4)_C \times U(2)_L \times U(2)_R]_{observable}\times
[U(2) \times USp(2) \times USp(10)]_{hidden}$, the
SM fermions and Higgs fields arise from
the intersections on the third two-torus,
and the complex structure parameters are
$6\chi_1= 2\chi_2= \chi_3= 2\sqrt{6}$.
To satisfy the RR tadpole cancellation conditions,
we choose $h_0m+3aq=-48$.}
\label{TI-U-4}
\end{center}
\end{table}




\begin{table}[h]

\begin{center}
\small
\begin{tabular}{|c||@{}c@{}||@{}c@{}|@{}c@{}|@{}c@{}|} \hline

Representation & Multiplicity &$U(1)_a$&$U(1)_b$&$U(1)_c$   \\
\hline \hline

$(4_a,\bar{2}_b)$ & 3 & 1 & -1 & 0    \\

$(\bar{4}_a,2_c)$ & 3 & -1 & 0 & 1     \\

$(2_b,\bar{2}_c)$ & 6 & 0 & 1 & -1     \\

\hline

$(4_a,2_c)$ & 3 & 1 & 0 & 1  \\

$(\bar{4}_a,\bar{2}_c)$ & 3 & -1 & 0 & -1   \\ \hline \hline

$6_a$ & 1 & 2 & 0 & 0  \\

$\overline{10}_a$ & 1 & -2 & 0 & 0  \\

$1_c$ & 2 & 0 & 0 & -2  \\

$3_c$ & 2 & 0 & 0 & 2 \\ \hline

\end{tabular}
\caption{The particle spectrum in observable sector in Model TIIA-I with
gauge symmetry $[U(4)_C\times U(2)_L\times
U(2)_R]_{observable} \times [U(2)\times USp(2)\times
USp(10)]_{hidden}$. Here,
$a$, $b$ and $c$ denote the
gauge groups $U(4)_C$, $U(2)_L$ and $U(2)_R$, respectively. }
\label{TI-U-4-spectrum-A}
\end{center}
\end{table}

\begin{table}[h]

\begin{center}
\small
\begin{tabular}{|c||@{}c@{}||@{}c@{}|@{}c@{}|@{}c@{}|@{}c@{}|} \hline

Representation & Multiplicity &$U(1)_a$&$U(1)_b$&$U(1)_c$& $U(1)_d$   \\
\hline \hline

$(4_a,\bar{2}_d)$ & 2 & 1 & 0 & 0 & -1    \\

$({\bar 4}_a,2_e)$ & 3 & -1 & 0 & 0 & 0 \\

$(4_a, 10_{O6})$ & 1 & 1 & 0 & 0 & 0    \\

$(\bar{2}_b,2_d)$ & 1 & 0 & -1 & 0 & 1    \\

$(\bar{2}_b,\bar{2}_d)$ & 5 & 0 & -1 & 0 & -1   \\

$(2_b,2_e)$ & 6 & 0 & 1 & 0 & 0    \\

$(2_c,2_d)$       & 2 & 0 & 0 & 1 & 1   \\

$({\bar 2}_c,10_{O6})$       & 2 & 0 & 0 & -1 & 0   \\

$(2_d, 2_e)$ & 6 & 0 & 0 & 0 & 1 \\ \hline

\end{tabular}
\caption{The exotic particle spectrum in Model TIIA-I with
gauge symmetry $[U(4)_C\times U(2)_L\times
U(2)_R]_{observable} \times [U(2)\times USp(2)\times
USp(10)]_{hidden}$. Here,
$a$, $b$, $c$, $d$, $e$ and $O6$ denote the
gauge groups $U(4)_C$, $U(2)_L$, $U(2)_R$, $U(2)$,
$USp(2)$ and $USp(10)$, respectively. }
\label{TI-U-4-spectrum-B}
\end{center}
\end{table}


Moreover, we can calculate the real parts of dilaton and complex structure moduli
\begin{eqnarray}
&&  {\rm Re} s = {{ 6^{3/4} e^{-\phi_4}}\over {24 \pi}}~,~
{\rm Re} u_1 = {{6^{3/4} e^{-\phi_4}}\over {2 \pi}}~,~
\nonumber\\&&
{\rm Re} u_2 ={{6^{3/4} e^{-\phi_4}}\over {6 \pi}}~,~
{\rm Re} u_3 = {{ 6^{3/4} e^{-\phi_4}}\over {12 \pi}}~.~\,
\label{Moduli}
\end{eqnarray}
And then, we obtain the gauge couplings at the string scale in
the observable sector
\begin{eqnarray}
2 g_{a}^{-2} =  g_{b}^{-2} =  g_{c}^{-2} = 5 {{ 6^{3/4} e^{-\phi_4}}\over {12 \pi}}~.~\,
\end{eqnarray}
So, the gauge couplings for $SU(2)_L$ and $SU(2)_R$ are unified at
the string scale, and are half of these for $SU(3)_C$ and $U(1)_{B-L}$.

Next, we shall study the flux parameter spaces
 when the fluxes $h_0$, $m$, $q$, and/or 
$a$ are very large. The general solution to $h_0m+3qa=-48$ is 
\begin{eqnarray}
m=2 N_1~,~~~h_0=-24N_2~,~~~q=2 N_3~,~~~a={{8(N_1N_2-1)}\over {N_3}}~,~\,
\end{eqnarray}
where $(N_1N_2-1)$ must be the multiple of $N_3$.
For example, we can have $N_3=\pm1$ or $N_3=\pm (N_1N_2-1)$.
If $a$ is very large, the theory will be strong coupled,
and the magnitude of the  cosmological constant is
very large. Although we can make the theory to be weakly coupled
and the magnitude of the  cosmological constant to be
small by introducing the very large fluxes $e$ and/or $e_0$,
we do not consider the  case with large $a$ for simplicity.
And then we obtain $N_3=\pm (N_1N_2-1)$.
To be concrete, we choose that $N_1$ and $N_2$ are positive integers, 
and $N_3= (N_1N_2-1)$. For simplicity, we assume $e_0=e=0$.

In the very large $N_1$ and/or $N_2$ limit, we obtain 
\begin{eqnarray}
\lambda_0=1-{{1}\over {N_1 N_2}} \longrightarrow 1~,~~~
\lambda \simeq {{10^{2/3}}\over {20}}~.~\,
\end{eqnarray}
In addition, we obtain
\begin{eqnarray}
{\rm Re}s \sim N_1 N_2^2~,~~~{\rm Re}t \sim N_2~,~~~
{\rm Re}u_i \sim N_1 N_2^2~,~\,
\end{eqnarray}
\begin{eqnarray}
e^{-\phi_4} \sim N_1 N_2^2~,~~~
e^{-\phi} \sim N_1 N_2^{1/2}~,~~
g_{a, b, c}^{-2} \sim N_1 N_2^2~,~\,
\end{eqnarray}
\begin{eqnarray}
V_0 \sim N_1^{-2} N_2^{-5}~.~\,
\end{eqnarray}
Thus, in the very large $N_1$ and/or $N_2$ limit, 
the theory  will be very weakly 
coupled,  the magnitude of
the cosmological constant will be close to zero,
and then the AdS vacua approach to the Minkowski vacua.

\subsection{Model TIIA-II}

In the above kind of models, we consider the scenario 
where the fluxes can be
very large while $(h_0m+3qa)$ is a negative constant. In this subsection,
we consider the scenario where the fluxes can be very large, and
 $(h_0m+3qa)$ is also a  large negative number. If  $(h_0m+3qa)$ is
negative and very large, we can alaways satisfy the RR tadpole
cancellation conditions by introducing enough D-branes. However,
we will have many chiral exotic particles that can not 
be decoupled easily. Thus, we
consider a special scenario where the complex structure parameters
$\chi_i$ are also very large, and we keep $h_0/(\chi_i \chi_j)$
and $a/(\chi_i \chi_j)$  constant. Then the very large
fluxes only affect the first RR tadpole cancellation condition
in Eq. (\ref{N-tadodhz2z2I}).

\begin{table}[h]
\begin{center}
\footnotesize
\begin{tabular}{|@{}c@{}|c||@{}c@{}c@{}c@{}||c|c||c|@{}c@{}|@{}c@{}|@{}c@{}||@{}c@{}|@{}c@{}|@{}c@{}|
@{}c@{}|} \hline

stk & $N$ & ($n_1$, $l_1$) & ($n_2$, $l_2$) & ($n_3$, $l_3$) & A &
S & $b$ & $b'$ & $c$ & $c'$ & $d$ & $e$ & $f$ & $O6$    \\
\hline \hline

$a$ & 4 & ( 1, 0) & (-1,-1) & (-1, 1) & 0 & 0 & 3 & 0(3) & -3 &
0(3) & 0(1) & 1 & -1 & 0(1)
\\ \hline

$b$ & 2 & (-1, 3) & (-2, 0) & ( 1, 1) & 0 & 0 & - & - & 6 & 0(1) &
-6 & 0(1) & 2 & 0(3)   \\ \hline

$c$ & 2 & ( 1, 3) & (-1, 1) & (-2, 0) & 0 & 0 & - & - & - & - & 6
& -2 & 0(1) & 0(3)
\\ \hline  \hline

$d$ & 1 & ( 1, 0) & ( 0,-2) & ( 0, 2) & 0 & 0 & - & - & - & - & -
& 0(2) & 0(2) & 0(4)
\\ \hline

$e$ & 3 & ( 0,-1) & ( 2, 0) & ( 0, 2) & 0 & 0 & - & - & - & - & -
& - & 0(4) & 0(2)
\\ \hline

$f$ & 3 & ( 0, 1) & ( 0, 2) & ( 2, 0) & 0 & 0 & - & - & - & - & -
& - & - & 0(2)
\\ \hline

$O6$ & $(1+\kappa)$ & ( 1, 0) & ( 2, 0) & ( 2, 0) & - & - &
\multicolumn{8}{|c|}{$\chi_3=\chi_2=6\chi_1$}
 \\ \hline
\end{tabular}
\caption{
D6-brane configurations and intersection numbers
for Model TIIA-II on Type IIA $\mathbf{T^6}$ orientifold.
The complete gauge symmetry is
$U(4)_C \times U(2)_L \times U(2)_R \times USp(2) \times
USp(6)^2 \times USp(2(1+\kappa))$, and the
SM fermions and Higgs fields arise from
the intersections on the first two-torus.} 
\label{TIIA-II}
\end{center}    
\end{table}

\begin{table}[h]

\begin{center}
\small
\begin{tabular}{|c||@{}c@{}||@{}c@{}|@{}c@{}|@{}c@{}|c|} \hline

Representation & Multiplicity &$U(1)_a$&$U(1)_b$&$U(1)_c$ & Field   \\
\hline \hline

$(4_a,\bar{2}_b)$ & 3 & 1 & -1 & 0  & $F_L(Q_L, L_L)$  \\

$(\bar{4}_a,2_c)$ & 3 & -1 & 0 & 1  & $F_R (Q_R, L_R)$   \\

$(2_b,\bar{2}_c)$ & 6 & 0 & 1 & -1  & $\Phi_i (H_u^i, H_d^i)$   \\

\hline

$(4_a, 2_c)$ & 3 & 1 & 0 & 1 &  \\

$(\bar{4}_a,\bar{2}_c)$ & 3 & -1 & 0 & -1  &  \\ \hline \hline

$(4_a, 6_e)$ & 1 & 1 & 0 & 0 & $X_{ae}$ \\

$(4_a, 6_f)$ & 1 & -1 & 0 & 0 & $X_{af}$ \\

$(2_b, 2_d)$ & 6 & 0 & -1 & 0 & $X^i_{bd}$ \\

$(2_b, 6_f)$ & 2 & 0 & 1 & 0 & $X^i_{bf}$ \\

$(2_c, 2_d)$ & 6 & 0 & 0 & 1 & $X^i_{cd}$ \\

$(2_c, 6_e)$ & 2 & 0 & 0 & -1 & $X^i_{ce}$ \\  \hline

\end{tabular}
\caption{The particle spectrum in observable sector in Model TIIA-II with
gauge symmetry $[U(4)_C\times U(2)_L\times
U(2)_R]_{observable} \times [ USp(2)\times USp(6)^2 \times 
USp(2+2 \kappa)]_{hidden}$. Here,
$a$, $b$ and $c$ denote the
gauge groups $U(4)_C$, $U(2)_L$ and $U(2)_R$, respectively. }
\label{TIIA-II-spectrum}
\end{center}
\end{table}

We present the D-brane configurations and 
intersection numbers in Table~\ref{TIIA-II}, and the particle spectrum in
the observed sector in Table~\ref{TIIA-II-spectrum}.
In addition, we choose
\begin{eqnarray}
\chi_1 = {{\sqrt \kappa} \over {3{\sqrt 2}}}~,~~
\chi_2=\chi_3= {\sqrt {2\kappa}}~,~\,
\label{chi-i}
\end{eqnarray}
\begin{eqnarray}
h_0 m + 3q a = -8\kappa~,~\,
\end{eqnarray}
where $\kappa=1,2,3,\cdots$. From Eqs. (\ref{ST-relation-IIA}),
(\ref{Eq-abchi}) and (\ref{Eq-hhchi}), we obtain the
generic solution to the flux consistent equations:
\begin{eqnarray}
 h_0 = 4 \kappa \eta~,~~ m=-4\eta ~,~~q=2 \eta'~,~~a=4\kappa \eta'~~,~\,
\end{eqnarray}
\begin{eqnarray}
b_1=6 \eta'~,~~b_2=b_3=36 \eta'~,~~h_1=-2 \eta~,~~h_2=h_3=-12 \eta~,~\,
\end{eqnarray}
where $\eta=\pm 1$ and $\eta'= \pm 1$.
In this case, $\kappa$ can be considered as a scale factor. 
Then in the very large $\kappa$ limit, we obtain 
\begin{eqnarray}
{\rm Re}s \longrightarrow \kappa^{-1} {\rm Re}s ~,~~
{\rm Re}t \longrightarrow {\rm Re}t~,~~
{\rm Re}u_i \longrightarrow {\rm Re} u_i~,~\,
\end{eqnarray}
\begin{eqnarray}
{\rm Im}s \longrightarrow {\rm Im}s~,~~
{\rm Im}t \longrightarrow {\rm Im}t~,~~
{\rm Im}u_i \longrightarrow {\rm Im} u_i~,~\,
\end{eqnarray}
\begin{eqnarray}
e^{-\phi_4} \longrightarrow \kappa^{-1/4} e^{-\phi_4}~,~~
e^{-\phi} \longrightarrow \kappa^{-1/4} e^{-\phi}~,~
V_0 \longrightarrow \kappa V_0~.~\,
\end{eqnarray}
Thus, the gauge theory will not become strong coupled
and the gauge couplings approach to the fixed constants. 
However, the magnitude of the
cosmological constant will be very large. In order to have 
 small magnitude for the cosmological constant,
we can introduce very large $e_0$ and/or $e$, as in the last Section.

Next, let us consider the phenomenological consequences.
The anomalies from three global $U(1)$s of 
$U(4)_C$, $U(2)_L$ and $U(2)_R$ 
are cancelled by the Green-Schwarz mechanism, and the gauge fields of 
these $U(1)$s obtain masses via the linear $B\wedge F$ couplings. So, the
effective gauge symmetry is $SU(4)_C\times SU(2)_L\times SU(2)_R$.
In order to break the gauge symmetry, on the first two-torus, 
we split the $a$ stack of
D-branes into $a_1$ and $a_2$ stacks with 3 and 1 D-branes,
respectively, and split the $c$ stack of D-branes into $c_1$ and
$c_2$ stacks with 1 D-brane for each one.
Then, the gauge symmetry is further broken down to 
$ SU(3)_C\times SU(2)_L\times U(1)_{I_{3R}}\times U(1)_{B-L}$.
We can break the $U(1)_{I_{3R}}\times U(1)_{B-L}$ gauge
symmetry down to the $U(1)_Y$ gauge symmetry by giving
vacuum expectation values
(VEVs) to the vector-like particles with quantum numbers
$({\bf { 1}, 1, 1/2, -1})$ and $({\bf { 1},
1, -1/2, 1})$ under $SU(3)_C\times SU(2)_L\times U(1)_{I_{3R}} \times
U(1)_{B-L} $ from $a_2 c_1'$ D-brane intersections. Similar to the
discussions in Ref.~\cite{Chen:2007px, Chen:2007zu}, we can explain 
the SM fermion masses and mixings  because all the SM fermions 
and Higgs fields arise 
from the intersections on the first two-torus.

To decouple the chiral exotic particles, we assume that 
the $e$ and $f$ stacks of the D-branes are on the top of
each other on the first two-torus,  the $d$ and $f$ stacks 
of the D-branes are on the top of each other on the 
second two-torus, and the $d$ and $e$ stacks of the D-branes 
are on the top of each other on the third two-torus.
Then we have the vector-like particles ($X_{ef}^i$ and 
$\overline{X}_{ef}^i$), ($X_{df}^j$ and 
$\overline{X}_{df}^j$), and ($X_{de}^j$ and 
$\overline{X}_{de}^j$) whose quantum numbers are $(6_e, 6_f)$,
$(2_d, 6_f)$, and $(2_d, 6_e)$, respectively, where $i=1, 2, 3, 4$
and $j=1, 2$. Then, we have the following
superpotential
\begin{eqnarray}
W & \supset & X_{ae} X_{af} (X_{ef}^i+ \overline{X}_{ef}^i)
+ X_{bd}^i X_{bf}^j (X_{df}^k+ \overline{X}_{df}^k)
+  X_{cd}^i X_{ce}^j (X_{de}^k+ \overline{X}_{de}^k)~,~
\end{eqnarray}
where we neglect the Yukawa couplings.
Thus, if  ($X_{ef}^i$ and 
$\overline{X}_{ef}^i$), ($X_{df}^j$ and 
$\overline{X}_{df}^j$), and ($X_{de}^j$ and 
$\overline{X}_{de}^j$) obtain VEVs, we might decouple
all the chiral exotic particles.
In short, below the string scale, we may just have the supersymmetric
Standard Model.

Moreover, we can calculate the real parts of the 
dilaton and complex structure moduli
\begin{eqnarray}
&&  {\rm Re} s = {{ {\sqrt {6 {\sqrt 2}}} e^{-\phi_4}}\over {4 \pi \kappa^{3/4}}}~,~
{\rm Re} u_1 = {{ {\sqrt {6 {\sqrt 2}}} \kappa^{1/4} e^{-\phi_4}}\over {2 \pi}}~,~
\nonumber\\&&
{\rm Re} u_2 =  {{ {\sqrt {6 {\sqrt 2}}} \kappa^{1/4} e^{-\phi_4}}\over {12 \pi}}~,~
{\rm Re} u_3 = {{ {\sqrt {6 {\sqrt 2}}} \kappa^{1/4} e^{-\phi_4}}\over {12 \pi}}~.~\,
\label{Moduli}
\end{eqnarray}
And then, we obtain the gauge couplings at the string scale in
the observable sector
\begin{eqnarray}
2 g_{a}^{-2} =  g_{b}^{-2} =  g_{c}^{-2} = 
{{ {\sqrt {6 {\sqrt 2}}} \kappa^{1/4} e^{-\phi_4}}\over {2 \pi}} 
\left({1\over {\kappa}}+{1\over 2}\right)~.~\,
\end{eqnarray}
So, the gauge couplings for $SU(2)_L$ and $SU(2)_R$ are unified at
the string scale, and are half of these for $SU(3)_C$ and $U(1)_{B-L}$.

\section{Flux Model Building on Type IIB orientifold}

We consider the Type IIB 
string theory compactified on a $\mathbf{T^6}$
orientifold where $\mathbf{T^{6}}$ is a six-torus
factorized as $\mathbf{T^{6}} = \mathbf{T^2} \times \mathbf{T^2}
\times \mathbf{T^2}$ whose complex coordinates are $z_i$, $i=1,\;
2,\; 3$ for the $i$-th two-torus, 
respectively~\cite{CU, BLT, Cvetic:2005bn}.
The orientifold projection is implemented by gauging the symmetry
$\Omega R$, where $\Omega$ is world-sheet parity, 
and $R$ is given by
\begin{eqnarray}
R: (z_1,z_2,z_3) \to (-z_1, -z_2, -z_3)~.~\,    \label{orientifold}
\end{eqnarray}
Thus, the model contains 64 O3-planes. 
In order to cancel the negative RR charges from these
O3-planes, we introduce the magnetized  
D(3+2n)-branes which are filling up the 
four-dimensional Minkowski space-time and wrapping
2n-cycles on the compact manifold. Concretely, for one stack
of $N_a$ D-branes wrapped $m_a^i$ times on the  $i$-th
two-torus $\mathbf{T^2_i}$, we turn on $n_a^i$ units of magnetic fluxes 
$F^i_a$ for the center of mass $U(1)_a$ gauge factor on $\mathbf{T^2_i}$, 
such that
\begin{eqnarray}
m_a^i \, \frac 1{2\pi}\, \int_{T^2_{\,i}} F_a^i \, = \, n_a^i ~,~\,
\label{monopole}
\end{eqnarray}
where $m_a^i$ can be half integer for tilted two-torus.
Then, the D9-, D7-, D5- and D3-branes contain 0, 1, 2 and 3 vanishing 
$m_a^i$s, respectively. Introducing for the $i$-th two-torus 
the even homology classes $[{\bf 0}_i]$ and $[{\bf T}^2_i]$ for 
the point and two-torus, respectively, the vectors of the RR 
charges of the $a$ stack of D-branes and its image are
\begin{eqnarray}
 && [{ \Pi}_a]\, =\, \prod_{i=1}^3\, ( n_a^i [{\bf 0}_i] + m_a^i [{\bf T}^2_i] ), 
 \nonumber\\&&
[{\Pi}_a']\, =\, \prod_{i=1}^3\, ( n_a^i [{\bf 0}_i]- m_a^i [{\bf T}^2_i] )~,~
\label{homology class for D-branes}
\end{eqnarray}
respectively.
The ``intersection numbers'' in Type IIA language, which determine 
the chiral massless spectrum, are
\begin{eqnarray}
I_{ab}&=&[\Pi_a]\cdot[\Pi_b]=\prod_{i=1}^3(n_a^im_b^i-n_b^im_a^i)~.~
\label{intersections}
\end{eqnarray}
Moreover, for a stack of $N$ D(2n+3)-branes whose homology classes
on $\mathbf{T^{6}}$ is (not) invariant under $\Omega R$, we obtain 
a ($U(N)$) $USp(2N)$  gauge symmetry with three (adjoint) 
anti-symmetric chiral superfields due to the orbifold projection.  
The physical spectrum is presented in Table \ref{spectrum}.

\begin{table}[t]
\caption{General spectrum  for magnetized D-branes on the  Type IIB
${\mathbf{T^6}}$ orientifold. }
\renewcommand{\arraystretch}{1.25}
\begin{center}
\begin{tabular}{|c|c|}
\hline {\bf Sector} & {\bf Representation}
 \\
\hline\hline
$aa$   & $U(N_a)$ vector multiplet  \\
       & 3 adjoint multiplets  \\
\hline
$ab+ba$   & $I_{ab}$ $(N_a,{\overline{N}}_b)$ multiplets  \\
\hline
$ab'+b'a$ & $I_{ab'}$ $(N_a,N_b)$ multiplets \\
\hline $aa'+a'a$ &$\frac 12 (I_{aa'} -  I_{aO3})\;\;$
symmetric multiplets \\
          & $\frac 12 (I_{aa'} +  I_{aO3}) \;\;$ 
anti-symmetric multiplets \\
\hline
\end{tabular}
\end{center}
\label{spectrum}
\end{table}

The flux models  on Type IIB orientifolds with 
four-dimensional $N=1$ supersymmetry  are primarily
constrained by the  RR tadpole cancellation conditions that
will be given later, the four-dimensional $N=1$ supersymmetry
condition, and the K-theory anomaly free conditions.
For the D-branes with world-volume magnetic field
$F_a^i={n_a^i}/({m_a^i\chi_i})$ where $\chi_i$ is the area of $\mathbf{T^2_i}$
in string units,  the condition to preserve the four-dimensional 
$N=1$ supersymmetry is~\cite{Cvetic:2005bn}
\begin{eqnarray}
\sum_i \left(\tan^{-1} (F_a^i)^{-1} + 
{\theta (n_a^i)} \pi \right)=0 ~~~{\rm mod}~ 2\pi~,~\,
\end{eqnarray}
where ${\theta (n_a^i)}=1$ for $n_a^i < 0$ and  
${\theta (n_a^i)}=0$ for $n_a^i \geq 0$.
The K-theory anomaly free conditions are
\begin{eqnarray}
&&  \sum_a N_a m_a^1 m_a^2 m_a^3 =  \sum_a N_a m_a^1 n_a^2 n_a^3
= \sum_a N_a n_a^1 m_a^2 n_a^3 
\nonumber\\&&
= \sum_a N_a n_a^1 n_a^2 m_a^3
=0 ~~~{\rm mod}~ 2~.~\,
\end{eqnarray}
And the holomorphic gauge kinetic function for a generic stack of
D(2n+3)-branes  is given by~\cite{Cremades:2002te, Lust:2004cx}
\begin{eqnarray}
f_a &=& {1\over {\kappa_a}}\left(  n_a^1\,n_a^2\,n_a^3\,s-
n_a^1\,m_a^2\,m_a^3\,t_1 \right.\nonumber\\&& \left.
-n_a^2\,m_a^1\,m_a^3\,t_2 -n_a^3\,m_a^1\,m_a^2\,t_3\right)~,~\,
\label{EQ-GKF}
\end{eqnarray}
where $\kappa_a$ is equal to  1 and 2 for $U(n)$ and $USp(2n)$,
respectively.



We turn on the NSNS flux $h_0$, RR flux $e_i$,  
non-geometric fluxes  $b_{ii}$ and ${\bar b}_{ii}$, and the
S-dual fluxes $f_i$, $g_{ij}$ and 
$g_{ii}$~\cite{Aldazabal:2006up, Chen:2007af}. To avoid the subtleties,
these fluxes should be even integers due to the Dirac quantization.
For simplicity, we assume 
\begin{eqnarray}
&& e_i=e~,~~b_{ii}=\beta~,~~ {\bar b}_{ii}={\bar \beta}~,~~
\nonumber\\&&
f_i=f~,~ g_{ij}=-g_{ii}=g~,~\,
\end{eqnarray}
where $i\not= j$.
Then the constraint on fluxes from Bianchi indetities is
\begin{eqnarray}
f {\bar \beta} ~=~g \beta~.~\,
\end{eqnarray}
The RR tadpole cancellation conditions are
\begin{eqnarray}
&& \sum_a N_an_a^1 n_a^2 n_a^3 =16 ~,~
\nonumber\\&&
\sum_a N_a n_a^i m_a^j m_a^k = -{1\over 2} e {\bar \beta}~,~
\nonumber\\&&
N_{{\rm NS7}_i}=0~,~~N_{{\rm I7}_i}=0~,~\,
\end{eqnarray}
where $i\not= j \not= k \not= i$, and the $N_{{\rm NS7}_i}$
and $N_{{\rm I7}_i}$ denote the NS 7-brane charge and
the other 7-brane charge, 
respectively~\cite{Aldazabal:2006up, Chen:2007af}.
Thus, if $e{\bar \beta} < 0$, the RR tadpole cancellation
conditions are relaxed elegantly because $-e{\bar \beta}/2$ only
needs to be even integer.
Moreover, we have 7 moduli fields in the supergravity
theory basis, the dilaton $s$, three K\"ahler moduli
$t_i$, and three complex structure moduli
$u_i$. With the above fluxes, we can assume 
\begin{eqnarray}
 t\equiv t_1+t_2+t_3~,~~~
u_1=u_2=u_3 \equiv u~.~\,
\end{eqnarray}
Then the superpotential becomes~\cite{Chen:2007af}
\begin{eqnarray}
{\cal W}=3 i e u + i h_0 s - t \left(\beta u - 
i {\bar \beta} u^2 \right) - s t \left(f- igu\right).~\,
\end{eqnarray}
The K\"ahler potential for these moduli is
\begin{eqnarray}
{\cal K} = -{\rm ln}(s+{\bar s})-\sum_{i=1}^3 {\rm ln} (t_i + {\bar t}_i)
-\sum_{i=1}^3 {\rm ln} (u_i + {\bar u}_i)~.~\,
\end{eqnarray}
Thus, for the supersymmetric Minkowski vacua, we have
\begin{eqnarray}
{\cal W}={\partial_s  {\cal W}} ={\partial_t  {\cal W}} 
={\partial_u  {\cal W}} =0 ~.~\,
\end{eqnarray}
From ${\partial_s  {\cal W}}={\partial_t  {\cal W}}=0$, 
we obtain~\cite{Chen:2007af} 
\begin{eqnarray}
t= {{ih_0}\over {f-igu}}~,~~~ s=-{{\beta}\over f} u~,~\,
\label{Moduli-Relations}
\end{eqnarray}
 then the superpotential turns out
\begin{eqnarray}
{\cal W} = \left( 3e -{{h_0 \beta}\over f}\right) i u~.~\,
\end{eqnarray}
Therefore, to satisfy ${\cal W}={\partial_u  {\cal W}} =0$, 
we obtain~\cite{Chen:2007af}
\begin{eqnarray}
3 e f = \beta h_0~.~\,
\end{eqnarray}
Because ${\rm Re} s > 0$,  ${\rm Re} t_i > 0$ and ${\rm Re} u_i > 0$,
we require 
\begin{eqnarray}
{{h_0}\over g} < 0~,~~~{{\beta}\over f} < 0~.~\,
\end{eqnarray}

In general, this kind of D-brane models might have the 
Freed-Witten anomalies~\cite{CU, Freed:1999vc}. 
Interestingly, the Freed-Witten anomalies can be cancelled by 
introducing additional D-branes~\cite{CU}. In particular, 
the additional D-branes will not affect the main properties of 
the D-brane models, for example, the four-dimensional $N=1$ 
supersymmetry and the chiral spectra, etc~\cite{CU}. 
Therefore, we can construct this kind of D-brane models by 
ignoring the subtlety of the Freed-Witten anomalies.



\section{Type IIB Flux Models}

Choosing $e {\bar \beta}=-12\kappa$ where $\kappa=1, 2, 3, ...$, 
we construct a series of realistic Pati-Salam models with 
gauge symmetry 
$U(4)_C \times U(2)_L \times U(2)_R \times USp(10) \times
USp(6(\kappa-1))^3$. We present their
D-brane configurations and intersection numbers 
in Table~\ref{TIIB-MI-Numbers}, and the resulting spectra in
Tables~\ref{Type-IIB-Spectrum-I} and \ref{Type-IIB-Spectrum-II}.
Table ~\ref{Type-IIB-Spectrum-I} is the spectra for $\kappa=1$
while Table~\ref{Type-IIB-Spectrum-II} is the additional
spectra due to $\kappa >1$.

 The anomalies from three global $U(1)$s of 
$U(4)_C$, $U(2)_L$ and $U(2)_R$ 
are cancelled by the Green-Schwarz mechanism, and the gauge fields of 
these $U(1)$s obtain masses via the linear $B\wedge F$ couplings. So, the
effective gauge symmetry is $SU(4)_C\times SU(2)_L\times SU(2)_R$.
In order to break the gauge symmetry, on the first two-torus, 
we split the $a$ stack of
D-branes into $a_1$ and $a_2$ stacks with 3 and 1 D-branes,
respectively, and split the $c$ stack of D-branes into $c_1$ and
$c_2$ stacks with 1 D-brane for each one.
Then, the gauge symmetry is further broken down to 
$ SU(3)_C\times SU(2)_L\times U(1)_{I_{3R}}\times U(1)_{B-L}$.
We can break the $U(1)_{I_{3R}}\times U(1)_{B-L}$ gauge
symmetry down to the $U(1)_Y$ gauge symmetry by giving
VEVs to the vector-like particles with quantum numbers
$({\bf { 1}, 1, 1/2, -1})$ and $({\bf { 1},
1, -1/2, 1})$ under $SU(3)_C\times SU(2)_L\times U(1)_{I_{3R}} \times
U(1)_{B-L} $ from $a_2 c_1'$ D-brane intersections. Similar to the
discussions in Ref.~\cite{Chen:2007px, Chen:2007zu}, 
we can explain the SM fermion masses and mixings via 
the Higgs fields $H_u^i$, $H_u'$, $H_d^i$ and
$H_d'$ because all the SM fermions and Higgs fields arise 
from the intersections on the first two-torus.

First, let us consider $\kappa=1$ case. Similar to the discussions 
in Ref.~\cite{Chen:2007af}, we can decouple the
chiral exotic particles in Table~\ref{Type-IIB-Spectrum-I}
as following.  We assume that 
the $T_R^i$ and $S_R^i$ obtain VEVs
at about the string scale, and their VEVs satisfy the 
D-flatness of $U(1)_R$. The chiral exotic particles can obtain
masses via the following superpotential
\begin{eqnarray}
W \supset {1\over {M_{\rm St}}} S_R^i S_R^j
 T_R^k T_R^l + T_R^i X^j X^k~,~\,
\end{eqnarray}
where $M_{\rm St}$ is the string scale, and
we neglect the order 1 (${\cal{O}}(1)$) coefficients. 
In addition, the vector-like particles $S_L^i$ and
$\overline{S}_L^i$ in the anti-symmetric 
representation of $SU(2)_L$  can obtain VEVs close to 
the string scale while keeping the D-flatness of $U(1)_L$. Thus, we can
decouple all the Higgs bidoublets close to the string scale  
except one pair of the linear combinations of the Higgs doublets 
for the electroweak symmetry breaking at the low
energy by fine-tuning the following superpotential
\begin{eqnarray}
W & \supset & \Phi_i ( \overline{S}_L^j \Phi' + S_R^j \overline{\Phi}')
+ \overline{\Phi}_i ( T_R^j \Phi' + S_L^j \overline{\Phi}')
\nonumber\\&&
+{1\over {M_{\rm St}}} \left( \overline{S}_L^i S_R^j \Phi_k \Phi_l +
S_L^i T_R^j \overline{\Phi}_k \overline{\Phi}_l 
\right.\nonumber\\&&\left.
+\overline{S}_L^i T_R^j \Phi' \Phi' 
+ S_L^i S_R^j \overline{\Phi}' \overline{\Phi}' \right)
~.~\,
\end{eqnarray}
In short, below the string scale for $\kappa=1$, we have the supersymmetric
SMs which may have zero, one, or a few SM singlets from $S_L^i$, 
$\overline{S}_L^i$, and/or $S_R^i$. And then the
low bound on the lightest CP-even Higgs boson mass in the
minimal supersymmetric SM can
be relaxed if we have the SM singlet(s) at low energy~\cite{Li:2006xb}.



\begin{table}[h]
\begin{center}
\footnotesize
\begin{tabular}{|@{}c@{}|c||@{}c@{}c@{}c@{}||c|c||c|@{}c@{}|@{}c@{}|@{}c@{}||
@{}c@{}|@{}c@{}|@{}c@{}|@{}c@{}|} \hline

stk & $N$ & ($n_1$, $l_1$) & ($n_2$, $l_2$) & ($n_3$, $l_3$) & A &
S & $b$ & $b'$ & $c$ & $c'$ & $d$ & $e$ & $f$ & $O3$    \\
\hline \hline

$a$ & 4 & ( 1, 0) & ( 1,-1) & ( 1, 1) & 0 & 0 & 3 & 0(3) & -3 &
0(3) & 0(1) & -1 & 1 & 0(1)
\\ \hline

$b$ & 2 & ( 1,-3) & ( 1, 1) & ( 1, 0) & 0 & 0 & - & - & 0(6) &
0(1) & -3 & 1 & 0(1) & 0(3)   \\ \hline

$c$ & 2 & ( 1, 3) & ( 1, 1) & ( 0,-1) & 6 & -6 & - & - & - & - &
0(3) & 0(1) & -1 & 3
\\ \hline  \hline

$d$ & $3(\kappa-1)$ & ( 1, 0) & ( 0,-2) & ( 0, 1) & 0 & 0 & - & -
& - & - & - & 0(2) & 0(1) & 0(2)
\\ \hline

$e$ & $3(\kappa-1)$ & ( 0,-1) & ( 2, 0) & ( 0, 1) & 0 & 0 & - & -
& - & - & - & - & 0(2) & 0(1)
\\ \hline

$f$ & $3(\kappa-1)$ & ( 0,-1) & ( 0, 2) & ( 1, 0) & 0 & 0 & - & -
& - & - & - & - & - & 0(2)
\\ \hline

$O3$ & 5 & ( 1, 0) & ( 2, 0) & ( 1, 0) & - & - &
\multicolumn{8}{|c|}{$2\chi_3=\chi_2=6\chi_1$}  \\ \hline

\end{tabular}
\caption{
D-brane configurations and intersection numbers
for a series of the models on Type IIB $\mathbf{T^6}$ orientifold.
The complete gauge symmetry is
$U(4)_C \times U(2)_L \times U(2)_R \times USp(10) \times
USp(6(\kappa-1))^3$, and the
SM fermions and Higgs fields arise from
the intersections on the first two-torus.
By the way, $l_{a}^{i}$ is equal to $m_{a}^{i}$ and $2m_{a}^{i}$
for  the rectangular and
tilted two-torus, respectively.}
\label{TIIB-MI-Numbers}
\end{center}    
\end{table}

\begin{table}[htb]
\footnotesize
\renewcommand{\arraystretch}{1.0}
\begin{center}
\begin{tabular}{|c||c||c|c|c||c|c|c|}\hline
 & Quantum Number
& $Q_4$ & $Q_{2L}$ & $Q_{2R}$  & Field \\
\hline\hline
$ab$ & $3 \times (4,\bar{2},1,1)$ & 1 & -1 & 0  & $F_L(Q_L, L_L)$\\
$ac$ & $3\times (\bar{4},1,2,1)$ & -1 & 0 & $1$   & $F_R(Q_R, L_R)$\\
$c_{S}$ & $6\times(1,1,\bar{3},1,1)$ & 0 & 0 & -2   & $T_R^i$  \\
$c_{A}$ & $6\times(1,1,1,1,1)$ & 0 & 0 & 2   & $S_R^i$ \\
$cO3$ & $3\times(1,1,2,10)$ & 0 & 0 & 1   &  $X^i$ \\
\hline\hline
$ac'$ & $3 \times (4,1,2,1)$ & 1 & 0 & 1  &  \\
& $3 \times (\bar{4}, 1, \bar{2},1)$ & -1 & 0 & -1 & \\
\hline
$bc$ & $6 \times (1,2,\overline{2},1)$ & 0 & 1 & -1   & 
$\Phi_i$($H_u^i$, $H_d^i$)\\
& $6 \times (1,\overline{2},2,1)$ & 0 & -1 & 1   & $\overline{\Phi}_i$ \\
\hline
$bc'$ & $1 \times (1,2, 2,1)$ & 0 & 1 & 1   & $\Phi'$($H'_u$, $H'_d$)\\
& $1 \times (1,\overline{2}, \overline{2},1)$ & 0 & -1 & -1   & 
$\overline{\Phi}'$ \\
\hline
$bb'$ 
 & $6 \times (1, 1, 1,1)$ & 0 & 2 & 0   & $S_L^i$ \\
 & $6 \times (1, \overline{1}, 1,1)$ & 0 & -2 & 0   & $\overline{S}_L^i$ \\
\hline
\end{tabular}
\caption{The chiral and vector-like superfields, and their quantum
numbers under the gauge symmetry $SU(4)_C\times SU(2)_L\times
SU(2)_R \times USp(10)$. This is the exact spectrum for $\kappa=1$.} 
\label{Type-IIB-Spectrum-I}
\end{center}
\end{table}


\begin{table}[h]

\begin{center}
\small
\begin{tabular}{|c||@{}c@{}||@{}c@{}|@{}c@{}|@{}c@{}|@{}c@{}|} \hline

Representation & Multiplicity &$U(1)_a$&$U(1)_b$&$U(1)_c$ & Field  \\
\hline \hline

$({\bar 4}_a, (6 (\kappa-1))_e)$ & 1 & -1 & 0 & 0 & $X_{ae}$ \\

$(4_a, (6 (\kappa-1))_f)$ & 1 & 1 & 0 & 0 & $X_{af}$    \\

$(\bar{2}_b, (6 (\kappa-1))_d))$ & 3 & 0 & -1 & 0 & $X^i_{bd}$   \\

$(2_b, (6 (\kappa-1))_e)$ & 1 & 0 & 1 & 0 &  $X_{be}$   \\

$({\bar 2}_c, (6 (\kappa-1))_f)$ &  1    & 0 & 0 & -1 & $X_{cf}$   \\ \hline

\end{tabular}
\caption{The chiral and vector-like superfields, and their quantum
numbers under the gauge symmetry $SU(4)_C\times SU(2)_L\times
SU(2)_R$ and $USp(6(\kappa-1))^3$. This is the additional spectrum due to
$\kappa > 1$.}
\label{Type-IIB-Spectrum-II}
\end{center}
\end{table}



Second, we consider the case with $\kappa > 1$, and decouple
the additional
chiral exotic superfields in Table~\ref{Type-IIB-Spectrum-II}.
We assume that the $e$ and $f$ stacks of the D-branes are on the 
top of each other on the first two-torus. Then we have the 
vector-like particles $X_{ef}^i$ and $\overline{X}_{ef}^i$
whose quantum numbers are $((6(\kappa-1))_e, (6(\kappa-1))_f)$ under 
the $USp(6(\kappa-1))_e \times USp(6(\kappa-1))_f$ gauge symmetry
 where $i=1,2$. Also, we assume that 
$X_{ef}^i$ and $\overline{X}_{ef}^i$ obtain VEVs.
Then, we have the following
superpotential
\begin{eqnarray}
W & \supset &  (X_{ef}^i+ \overline{X}_{ef}^i) X_{ae} X_{af}
+ S_L^i X_{bd}^j X_{bd}^j + \overline{S}_L^i X_{be} X_{be}
+S_R^i X_{cf} X_{cf}~.~\,
\end{eqnarray}
Thus, we can  decouple
 the extra chiral exotic particles due to $\kappa > 1$ as well.
In short, below the string scale, we also have the supersymmetric
SMs with or without additional SM singlets.



Next, we consider the gauge coupling unification and moduli stabilization.
The real parts of the dilaton and K\"ahler moduli in our model 
are~\cite{Chen:2007af}
\begin{eqnarray}
&&  {\rm Re} s = {{{\sqrt 6} e^{-\phi_4}}\over {4 \pi}}~,~
{\rm Re} t_1 = {{{\sqrt 6} e^{-\phi_4}}\over {2 \pi}}~,~
\nonumber\\&&
{\rm Re} t_2 ={{{\sqrt 6} e^{-\phi_4}}\over {12 \pi}}~,~
{\rm Re} t_3 = {{{\sqrt 6} e^{-\phi_4}}\over {6 \pi}}~,~\,
\label{Moduli}
\end{eqnarray}
where $\phi_4$ is the four-dimensional dilaton. From Eq. (\ref{EQ-GKF}),
 we obtain that the SM gauge couplings are unified at the string
scale as follows
\begin{eqnarray}
g_{SU(3)_C}^{-2} = g_{SU(2)_L}^{-2} ={3\over 5} g_{U(1)_Y}^{-2}
={{{\sqrt 6} e^{-\phi_4}}\over {2 \pi}}~.~\,
\end{eqnarray}
Using the unified gauge coupling $g^{2} \simeq 0.513$
in supersymmetric SMs at the GUT scale, we get
\begin{eqnarray}
 \phi_4 \simeq - 1.61~.~\,
\end{eqnarray}

For moduli stabilization, we first obtain $t$
from Eqs. (\ref{Moduli-Relations}) and (\ref{Moduli})
\begin{eqnarray}
{\rm Re} t = {{3{\sqrt 6} e^{-\phi_4}}\over {4 \pi}}
~,~ {\rm Im} t= \pm {\sqrt {{{3 \beta h_0}\over {fg}}- 
 {{27 e^{-2\phi_4}}\over {8 \pi^2}}}}  ~.~\,
\end{eqnarray}
Thus, we have
\begin{eqnarray}
&& {\rm Im} s = -{1\over 3} {\rm Im} t +{{\beta}\over g}~,~
\nonumber\\&&
{\rm Re} u = - {{{\sqrt 6} f e^{-\phi_4}}\over {4 \pi \beta}}~,~~
{\rm Im} u = {f\over {3\beta}} {\rm Im} t -{{f}\over g}~.~
\end{eqnarray}



In this paper, we shall choose the
unified gauge coupling at the string scale as the unified 
gauge coupling in the supersymmetric SM at
the GUT scale, in other words, we choose
$\phi_4=-1.61$. And then  we obtain
\begin{eqnarray}
{\rm Re} s= 0.975~,~~{\rm Re} t_1=1.95~,~~
{\rm Re} t_2=0.325~,~~{\rm Re} t_3=0.650~.~\,
\end{eqnarray}
In the following discussions, we always have the above values 
for ${\rm Re} s$, ${\rm Re} t_1$, ${\rm Re} t_2$, and ${\rm Re} t_3$. 
So, we will not present them again.
Also, we will assume that 
$k_0$, $k_1$, $k_2$ and $k_3$ are positive integers, and
\begin{eqnarray}
 k_0 k_1=3~,~~~\eta=\pm 1~,~~~\eta'=\pm 1~.~\,
\end{eqnarray}



\subsection{Models with  $\kappa=1$}

First, we consider the models with $\kappa=1$. Although this kind
of models is the same as that in Ref.~\cite{Chen:2007af}, 
we consider the generic solution to the flux
consistent equations, and find a huge number of flux models.
The general solution to the flux consistent equations is
\begin{eqnarray}
&& e=2 k_0 \eta~,~~{\bar \beta}=-2 k_1 \eta~,~~f=-2 k_2\eta'~,~
\nonumber\\&&
\beta=2 k_3 \eta'~,~~ g={{2 k_1 k_2 \eta} \over {k_3}}~,~~
h_0=-{{6k_0 k_2 \eta} \over {k_3}}~,~\,
\end{eqnarray}
where  $k_1 k_2$ and $3k_0 k_2$ are  multiples of $k_3$.

There are three kinds of possible solutions that can satisfy 
$e{\bar \beta}=-12$, so, let us discuss them one by one
in the following: \\

(1) For $k_0$, $k_1$, $k_2$ and $k_3$, we have
\begin{eqnarray}
k_0=3~,~~k_1=1~,~~k_2=nN~,~~k_3=N~.~\,
\end{eqnarray}
So we obtain
\begin{eqnarray}
&& e=6 \eta~,~~{\bar \beta}=-2 \eta~,~~f=-2 n N \eta'~,~
\nonumber\\&&
\beta=2 N \eta'~,~~ g= 2 n \eta~,~~
h_0=-18 n \eta~.~\,
\end{eqnarray}
Choosing $\phi_4=-1.61$, we obtain 
\begin{eqnarray}
&& {\rm Re} u = 0.975 n~,~~
{\rm Im} t \equiv
\sum_{i=1}^3 {\rm Im} t_i = \pm {\sqrt {{27\over n}-8.56}}~,~~~
\nonumber\\&&
{\rm Im} s = -{1\over 3} {\rm Im} t +{{N\eta'}\over {n \eta}}~,~~
{\rm Im} u = -{n\over 3} {\rm Im} t +{{N\eta'}\over { \eta}} ~.~\,
\end{eqnarray}
Thus, in order to have $({\rm Im} t)^2 > 0$, we obtain $n < 4$.
And in the very large $N$ limit, we obtain that only ${\rm Im} s$ and
${\rm Im} u$ will become very large as follows
\begin{eqnarray}
{\rm Im} s \sim N~,~~~ {\rm Im} u \sim N ~.~\,
\end{eqnarray} \\

(2) For $k_0$, $k_1$, $k_2$ and $k_3$, we have
\begin{eqnarray}
k_0=1~,~~k_1=3~,~~k_2=nN~,~~k_3=N~.~\,
\end{eqnarray}
So we obtain
\begin{eqnarray}
&& e=2 \eta~,~~{\bar \beta}=-6 \eta~,~~f=-2 n N \eta'~,~
\nonumber\\&&
\beta=2 N \eta'~,~~ g= 6 n \eta~,~~
h_0=-6 n \eta~.~\,
\end{eqnarray}
With $\phi_4=-1.61$, we find $({\rm Im} t)^2 < 0$. So,
this solution is not a correct solution. \\

(3) For $k_0$, $k_1$, $k_2$ and $k_3$, we have
\begin{eqnarray}
k_0=1~,~~k_1=3~,~~k_2=nN~,~~k_3=3N~.~\,
\end{eqnarray}
So we obtain
\begin{eqnarray}
&& e=2 \eta~,~~{\bar \beta}=-6 \eta~,~~f=-2 n N \eta'~,~
\nonumber\\&&
\beta=6 N \eta'~,~~ g= 2 n \eta~,~~
h_0=-2 n \eta~.~\,
\end{eqnarray}
With $\phi_4=-1.61$, we obtain $n=1$ from
$({\rm Im} t)^2 > 0$.  And we have 
\begin{eqnarray}
&& {\rm Re} u = 0.325~,~~
{\rm Im} t = \pm 0.664~,~~~
\nonumber\\&&
{\rm Im} s = \mp 0.221 +{{3 N\eta'}\over { \eta}}~,~~
{\rm Im} u = \mp 0.0738 +{{N\eta'}\over { \eta}} ~.~\,
\end{eqnarray}
In the very large $N$ limit, we obtain that only ${\rm Im} s$ and
${\rm Im} u$ will become very large as follows
\begin{eqnarray}
{\rm Im} s \sim N~,~~~ {\rm Im} u \sim N ~.~\,
\end{eqnarray}

\subsection{Infinity Flux Vacua for  $\kappa > 1$}

We consider the models with $\kappa > 1$. Comparing to the above
models, we have additional gauge 
symmetry $USp(6(\kappa-1))^3$. There are four kinds of solutions
to the flux consistent equations, and we will study them in
the following: \\

(1)  The first kind of solutions to the flux consistent equations is
\begin{eqnarray}
&& e=2 k_0 \kappa \eta~,~~{\bar \beta}=-2 k_1 \eta~,~~f=-2 k_2\eta'~,~
\nonumber\\&&
\beta=2 k_3 \eta'~,~~ g={{2 k_1 k_2 \eta} \over {k_3}}~,~~
h_0=-{{6k_0 k_2 \kappa \eta} \over {k_3}}~,~\,
\end{eqnarray}
where $k_1 k_2$ and $3 k_0 k_2 \kappa $ are the multiples of $k_3$.
In order to have positive $({\rm Im} t)^2$, we obtain 
\begin{eqnarray}
{{9 k_0 k_3 \kappa}\over {k_1 k_2}} ~\geq~ 8.56 ~.~\,
\end{eqnarray}
Thus, there exists a huge number of solutions to the flux consistent
equations.

Let us consider a special case with finite $k_i$ while very 
large $\kappa$, {\it i.e.}, only $\kappa$ can be very large.
In the very large $\kappa$ limit, we obtain that only ${\rm Im} s$,  
${\rm Im} t$, 
and ${\rm Im} u$ will become very large as follows
\begin{eqnarray}
{\rm Im} s \sim {\sqrt {\kappa}}~,~~ {\rm Im} t \sim {\sqrt {\kappa}}~,~~
 {\rm Im} u \sim {\sqrt {\kappa}}~.~\,
\end{eqnarray} 
In this case, let us present a concrete example. We choose
\begin{eqnarray}
&& e= 6 \kappa \eta~,~~{\bar \beta}=-2 \eta~,~~f=-2 \eta'~,~
\nonumber\\&&
\beta=2  \eta'~,~~ g=2 \eta~,~~
h_0=-18 \kappa \eta~.~\,
\end{eqnarray}
With $\phi_4=-1.61$,  we have 
\begin{eqnarray}
&&  {\rm Re} u = 0.975~,~~~
{\rm Im} t = \pm {\sqrt {27 \kappa -8.56}}~,~\,
\nonumber\\&&
{\rm Im} s = {\rm Im} u = -{1\over 3} {\rm Im} t + {{\eta'}\over {\eta}} ~.~\,
\end{eqnarray} \\

(2)  The second kind of solutions to the flux consistent equations is
\begin{eqnarray}
&& e=2 k_0  \eta~,~~{\bar \beta}=-2 k_1 \kappa \eta~,~~f=-2 k_2\eta'~,~
\nonumber\\&&
\beta=2 k_3 \eta'~,~~ g={{2 k_1 k_2 \kappa \eta} \over {k_3}}~,~~
h_0=-{{6k_0 k_2 \eta} \over {k_3}}~,~\,
\end{eqnarray}
where $k_1 k_2 \kappa$ and $3k_0 k_2  $ are the multiples of $k_3$.
In order to have positive $({\rm Im} t)^2$, we obtain 
\begin{eqnarray}
{{9 k_0 k_3}\over {k_1 k_2 \kappa}} ~\geq~ 8.56 ~.~\,
\end{eqnarray}
And then, we obtain 
\begin{eqnarray}
k_3~=~k_1 k_2 \kappa~,~~~{\rm for }~~k_0=1~,~\,
\end{eqnarray}
\begin{eqnarray}
k_3~=~k_1 k_2 \kappa/n~,~~~{\rm for }~~k_0=3~,~\,
\end{eqnarray}
where $n=1, 2, 3$.
Thus, there also exists a huge number of solutions to the flux consistent
equations.

However, for very large $\kappa$ and finite $k_i$, 
we obtain that $({\rm Im} t)^2$ is
negative, and then we do not have such kind of solutions. \\

(3) The third kind of solutions to the flux consistent equations is
\begin{eqnarray}
&& e=2 k_0 N \eta~,~~{\bar \beta}=-2 k_1 N \eta~,~~f=-2 k_2 N \eta'~,~
\nonumber\\&&
\beta=2 k_3 N \eta'~,~~ g={{2 k_1 k_2 N \eta} \over {k_3}}~,~~
h_0=-{{6k_0 k_2 N \eta} \over {k_3}}~,~\,
\end{eqnarray}
where $N$ is a positive integer, and
$k_1 k_2 N$ and $3k_0 k_2 N $ are the multiples of $k_3$.
Also, we have
\begin{eqnarray}
\kappa~=~ N^2~.~\,
\end{eqnarray}
In order to have positive $({\rm Im} t)^2$, we obtain 
\begin{eqnarray}
{{9 k_0 k_3 }\over {k_1 k_2}} ~\geq~ 8.56 ~.~\,
\end{eqnarray}
Thus, there still exists a huge number of solutions to the flux consistent
equations. In fact, we just rescale all the
fluxes by $N$. Note that the flux ratios are independent on $N$,
we obtain that ${\rm Re} u$, ${\rm Im} s$, ${\rm Im} t$
and ${\rm Im} u$ are constants since they only depend 
on the flux ratios.  In this case, all the phenomenological
discussions are similar to those in Ref.~\cite{Chen:2007af} except
that we have additional $USp(6\kappa-6)^3$ gauge symmetry 
and the corresponding extra 
chiral exotic particles in Table~\ref{Type-IIB-Spectrum-II}.

Let us give a concrete example for finite $k_i$ and very large $N$. 
We choose
\begin{eqnarray}
&& e= 6 N \eta~,~~{\bar \beta}=-2 N \eta~,~~f=-2 N\eta'~,~
\nonumber\\&&
\beta=2 N  \eta'~,~~ g=2 N \eta~,~~
h_0=-18 N \eta~.~\,
\end{eqnarray}
With $\phi_4=-1.61$,  we have 
\begin{eqnarray}
{\rm Re} u = 0.975~,~~ {\rm Im} t = \pm 4.30~,~~
{\rm Im} s = {\rm Im} u = \mp 1.43 +1 ~.~\,
\end{eqnarray}
This is similar to the example in Ref.~\cite{Chen:2007af}. \\

(4)  The fourth kind of solutions to the flux consistent equations is
\begin{eqnarray}
&& e=2 k_0 N \eta~,~~{\bar \beta}=-2 k_1 N \eta~,~~f=-2 k_2\eta'~,~
\nonumber\\&&
\beta=2 k_3 N \eta'~,~~ g={{2 k_1 k_2  \eta} \over {k_3}}~,~~
h_0=-{{6k_0 k_2 \eta} \over {k_3}}~,~\,
\end{eqnarray}
where $k_1 k_2 $ and $3k_0 k_2 $ are the multiples of $k_3$.
Also, we have $\kappa = N^2$. In order to have positive 
$({\rm Im} t)^2$, we obtain 
\begin{eqnarray}
{{9 k_0 k_3 N }\over {k_1 k_2}} ~\geq~ 8.56 ~.~\,
\end{eqnarray}
So, we also have a huge number of solutions to the flux consistent
equations.

If $N$ is very large while $k_i$ is finite,
we obtain that  ${\rm Re} u$, ${\rm Im} s$,  
${\rm Im} t$, and ${\rm Im} u$ will depend on $N$ as follows
\begin{eqnarray}
{\rm Re} u \sim {1\over N} \sim {1\over {\sqrt \kappa}}~,~~
{\rm Im} s \sim N \sim {\sqrt \kappa}~,~~ 
{\rm Im} t \sim {\sqrt {N}} \sim  \kappa^{1/4}~,~~
{\rm Im} u \sim {1\over {\sqrt {N}}} \sim {1\over {\kappa^{1/4}}}
~.~\,
\end{eqnarray} 
Thus, in the very large $N$ limit, ${\rm Re} u$ and ${\rm Im} u$ 
will be very small while
${\rm Im} s$ and ${\rm Im} t$ will be very large.
Let us present a concrete example for this case. We choose
\begin{eqnarray}
&& e= 6 N \eta~,~~{\bar \beta}=-2 N \eta~,~~f=-2 \eta'~,~
\nonumber\\&&
\beta=2 N \eta'~,~~ g=2 \eta~,~~
h_0=-18 \eta~.~\,
\end{eqnarray}

With $\phi_4=-1.61$,  we have 
\begin{eqnarray}
&& {\rm Re} u = {{0.975} \over N}~,~~
{\rm Im} t = \pm {\sqrt {27N -8.56}}~,~~~
\nonumber\\&&
{\rm Im} s =  -{1\over 3} {\rm Im} t + {{N\eta'}\over {\eta}}~,~~~
{\rm Im} u =  -{1\over {3N}} {\rm Im} t + {{\eta'}\over {\eta}}~.~\,
\end{eqnarray}

\section{Discussion and Conclusions}

In this paper, we studied the detail flux parameter spaces
 for semi-realistic supersymmetric Pati-Salam 
models in the AdS vacua on Type IIA orientifold 
and realistic supersymmetric Pati-Salam models 
 in the Minkowski vacua on Type IIB orientifold
with general flux compactifications. We have 
shown that there indeed exists a huge number of semi-realistic 
Type IIA  and realistic Type IIB flux models. 
So, we do have the string landscape for the flux models.
However, these semi-realistic Tyep IIA and realistic Type IIB 
flux models can not be populated in the string landscape.

For the supersymmetric intersecting D6-brane model building 
in the AdS vacua on Type IIA orientifold with flux compactifications,
if we keep $(e_0a-eh_0)$ as a constant
while allowing  $e$ and/or $e_0$ to be very large, the very large
fluxes do not affect the main properties of the models, 
and only ($ 3a {\rm Im} s +\sum_{i=1}^3 b_i {\rm Im} u_i$)
will be proportional to $e$ and then very large.
Also, if $(e_0a-eh_0)$
is very large, we showed that the theory is very weakly coupled, and 
the magnitude of the cosmological constant becomes very small. 
And if only $e_0$ is very large, 
($ 3a {\rm Im} s +\sum_{i=1}^3 b_i {\rm Im} u_i$) will be a constant
as well since it does not depend on $e_0$. Moreover, 
we considered two kinds of semi-realistic Pati-Salam models with 
very large fluxes $a$, $h_0$, $m$, and/or $q$. In the first kind of
models, we keep $(h_0m +3qa)$ as a negative constant. So we do not change 
the RR tadpole cancellation conditions, the gauge symmetries, 
and the particle spectra
of the models due to the very large fluxes. In the very large flux limit,
the theory becomes very weakly coupled, and the magnitude of the
cosmological constant will become very small as well.
In the second kind of Pati-Salam models, 
we took not only  $a$ and $h_0$ to be very large, but also 
$(h_0m +3qa)$ to be negative and very large. In particular, 
we took the complex structure moduli to be very large so that 
only one of the RR tadpole cancellation conditions
is very large and proportional to $a$ or $h_0$.
We showed that the gauge symmetry can be broken down to the SM gauge
symmetry, and the exotic particles might be decoupled.
The gauge couplings for $SU(2)_L$ and $SU(2)_R$ are half of these
for $SU(3)_C$ and $U(1)_{B-L}$. In the very large flux limit,
the gauge coulings approach to the fixed constants. However, 
the magnitude of the cosmological constant  will be very large, which
can become very small again if we introduce very large fluxes
$e$ and/or $e_0$.

For the supersymmetric D-brane model building in the Minkowski vacua 
on Type IIB orientifold with general flux compactifications,
we constructed a series of realistic Pati-Salam models with  gauge group
$U(4)_C \times U(2)_L \times U(2)_R \times USp(10) \times USp(6(\kappa-1))^3$.
Interestingly, in the very large flux limit, we can choose 
the unified gauge coupling at the string scale as the unified gauge 
coupling in the supersymmetric SMs at the GUT scale
because the real parts of the dilaton  and K\"ahler moduli can 
be independent of the very large fluxes.  For the first kind of models with
 $\kappa=1$,  the very large fluxes do not contribute to the RR tadpole
cancellation conditions, and we found two kinds of solutions to 
the flux consistent equations. 
In the very large flux limit, the real part of the complex 
structure moduli  and the sum of the imaginary 
parts of the K\"ahler moduli 
are constants and independent of the very large fluxes. Only the imaginary
parts of the dilaton and complex structure moduli become very large
and proportional to the very large fluxes.
For the second kind of models with $\kappa > 1$, we obtained four kinds of 
solutions to the flux consistent equations, and showed that there
indeed exists a huge number of the realistic flux models.
In the very large $\kappa$ limit, we consider the asymptotic behaviour
 for ${\rm Re} u$, ${\rm Im}s$, ${\rm Im}t$, and ${\rm Im}u$,
and present some concrete examples.

\section*{Acknowledgments}

 This research was supported in part by
the Mitchell-Heep Chair in High Energy Physics (CMC), by the
Cambridge-Mitchell Collaboration in Theoretical Cosmology (TL),
and by the DOE grant DE-FG03-95-Er-40917 (DVN).

\end{document}